\def\lae{\mathrel{<\kern-1.0em\lower0.9ex\hbox{$\sim$}}}
\pdfoutput=1
\documentclass[12pt,preprint]{aastex}

\newcommand{\gae}{\mathrel{>\kern-1.0em\lower0.9ex\hbox{$\sim$}}}

\lefthead{Boroson et al.}
\righthead{One Her X-1 Orbit with {\it FUSE}}

\begin{document}

\title{FUSE Observations of a Full Orbit of Hercules X-1: Signatures of 
Disk, Star, and Wind$^1$}

\author{Bram S. Boroson}
\affil{bram.boroson@gmail.com}

\and

\author{Saeqa Dil Vrtilek and John C. Raymond}
\affil{Smithsonian Astrophysical Observatory
Mail Stop 83
Cambridge, MA 02138
saku@head.cfa.harvard.edu, jraymond@cfa.harvard.edu}

\and

\author{Martin Still}
\affil{South African Astronomical Observatory
PO Box 9
Observatory 7935
Cape Town, South Africa
still@saao.ac.za}

$^1$Based on observations made with the NASA-CNES-CSA Far Ultraviolet 
Spectroscopic Explorer. FUSE is operated for NASA by the Johns Hopkins 
University under NASA contract NAS5-32985


\begin{abstract}

We observed an entire 1.7 day orbit of the X-ray binary Hercules~X-1 with
the Far Ultraviolet Spectroscopic Explorer ({\it FUSE}).  Changes in the
O\,{\sc vi}$\lambda\lambda1032,1037$ line profiles through eclipse 
ingress and
egress indicate a Keplerian accretion disk spinning prograde with the
orbit. These observations may show the first double-peaked accretion disk
line profile to be seen in the Hercules~X-1 system.  Doppler tomograms of
the emission lines show a bright spot offset from the Roche lobe of the
companion star HZ~Her, but no obvious signs of the accretion disk. 
Simulations show that the bright spot is too far offset from the Roche
lobe to result from uneven X-ray heating of its surface. The absence of
disk signatures in the tomogram can be reproduced in simulations which 
include absorption from a stellar wind. We attempt to diagnose
the state of the emitting gas from the C\,{\sc iii}$\lambda977$,
C\,{\sc iii}$\lambda1175$, and N\,{\sc III}$\lambda991$ emission lines. The 
latter may be enhanced through Bowen fluorescence. 
\end{abstract}


\section{Introduction}

Hercules X-1 (discovered by Tananbaum et al. 1972) is one of the most
frequently observed X-ray binary systems. The intermediate mass of the
donor star, $\approx 2.2$~M$_\odot$, leads to a wealth of behavior seen in
both low mass and high mass systems. Although we suspect the mass flow is
primarily through Roche lobe overflow giving rise to an accretion disk, as
in the Low Mass X-ray Binaries (LMXB), the variable P~Cygni profiles
observed in UV lines suggest either a transient stellar wind or a stellar
wind that is photoionized in some regions (Boroson, Kallman, \&\ Vrtilek
2001). There are theoretical grounds as well to expect that winds may
arise in this system when X-rays from the neutron star heat the surface of
the normal star or accretion disk (Arons 1973; Davidson \&\ Ostriker 1973; 
Basko et al. 1977; London, McCray, \&\ Auer 1981; Begelman, McKee, \&\
Shields 1983; Begelman \&\ McKee 1983). 

Most LMXBs, including Z-sources and atoll-sources, do not show 
persistent pulsations, perhaps because they have neutron 
stars with low magnetic fields.  
These sources thus lack an essential window into their kinematics. 
Light travel-time delays in the 1.24~second Her~X-1 X-ray pulsation period
determine that the neutron star is in a nearly circular 1.7~day orbit with
semimajor axis a$_{\rm x}=13.86$~seconds. The orbital period is slowly
lengthening from mass loss in the system (Deeter et al. 1991). 

Fortuitously, Her~X-1 has high orbital inclination ($>80$ degrees),
allowing total X-ray eclipses to constrain the size of the donor star, and
the high galactic latitude offers only small reddening, allowing the
system to be observed at crucial UV and even EUV wavelengths. 

The system's most mysterious variability is the 35-day X-ray high and low
cycle (Giacconi et al. 1973). For 8--11 days (the ``Main-On state'') of
this cycle Her~X-1 emits X-rays with a $\sim 10^{37}$ erg~s$^{-1}$ total
X-ray luminosity. For a $\sim 4$ day cycle (the ``Short-On state''),
halfway through the 35 day cycle, Her~X-1's output is several times lower. 
For the remaining portion of the 35 days, the Off state, the observed
X-ray flux is only several percent of that during the Main-On state. 

The behavior of the optical flux and X-ray spectrum over the 35-day period
shows that the 35-day variability is not isotropic. Instead, the
X-ray emission is merely obscured by a warped accretion disk whose shape
``precesses" globally throughout a 35-day period. The optical 
emission from the normal star varies with the 1.7~day binary orbit.
Its spectral type is variously classified as A through F, because X-rays 
from the neutron star heat the surface of the facing side of the Roche 
lobe. The spectral type continues to change with the orbital period 
throughout the longer 35-day
cycle, implying that X-rays continue to heat the surface of HZ~Her.

The transition between the Off and Main-On state is only a 
few hours, whereas the Short-On state is entered into more gradually.
This indicates that regions of the disk at different radii may obscure 
the central source, in which case the disk must precess globally with the 
same period. The variation in pulse shape with the long-term period is 
sometimes taken to imply that the innermost edge of the accretion disk 
precesses with the 35 day period, obscuring local regions on the neutron 
star (Scott, Leahy, \&\ Wilson, 2000).

The physical cause of a 35-day global disk precessing disk warp, or a
definitive relation between system parameters and disk precession period,
remains unknown and is a major goal of investigations of the system.  It
has been proposed, but not confirmed, that reprocessed X-ray radiation
pressure (Maloney \&\ Begelman 1997; Wijers \&\ Pringle 1999) or an X-ray
driven disk wind and corona (Meyer, Meyer-Hofmeister, 1984) could
propagate the warped shape in the disk. Similar "long-term periods'' have
been observed in other X-ray binaries, including LMC~X-4.  Her~X-1 serves
as a prototype of this behavior. 

X-ray dips present another variability that has not been explained 
definitively.  Crosa \&\ Boynton (1980) showed that the dips recur
every 1.65 days, near, but not at, the 1.62~day beat period between
orbital and precessional periods. This was confirmed by long-term X-ray 
observations (Scott \&\ Leahy, 1999). The dips are thought to be associated 
with the gas stream between the stars, but there is no consensus 
explanation for them or their period.

As befits such a frequently observed, enigmatic, yet prototypical source,
the methods of spectroscopy have been applied to the system at a wide
range of wavelengths. These studies have borne fruit with determinations
of the system's elemental abundances and the orbital motion of the normal
star (the neutron star's pulse delays indicate only the motion of one
component of the binary). 

High resolution spectroscopy in the X-ray range using {\it 
XMM} (Jimenez-Garate et al. 2002) shows a
multitude of X-ray emission lines presumably from an accretion disk
corona. The line ratios indicate that the gas is enhanced from CNO
processing from a massive progenitor.  Jimenez-Garate et al. (2005)
confirmed the CNO enhancement by observations of more than two dozen
emission lines originating in an accretion disk corona. Model predictions
of the disk corona's response to illumination by the central X-ray source
are in reasonable agreement with the observed fluxes for low and moderate
Z elements ( O through S), but the Fe\,{\sc xxv},Fe\,{\sc xxvi} lines are 
several times brighter than predicted (Jimenez-Garate et al. 2005). 

From observations of optical absorption lines (Reynolds et al. 1997), and 
from optical pulsations that result from the X-rays from the neutron star 
periodically striking the surface of the normal star (Middleditch \&\ 
Nelson 1976), we know that the normal star has a mass $\approx2.2$~M$_\odot$
and the neutron star has a mass 1.5$\pm0.3$~M$_\odot$.
The optical spectrum shows absorption lines from HZ~Her, emission at the 
Bowen blend near 4640\AA\ (see Schachter, Filippenko, \&\ Kahn 1989 for 
a discussion of the formation of these lines), 
and emission from He\,{\sc ii}$\lambda4686$. Still et al. (1997) performed 
Doppler tomography on these optical emission lines but could not 
conclude that they arose in a symmetric accretion disk.

The UV spectrum is crucial for an understanding of the accretion disk and
the X-ray illuminated face of HZ~Her. The continuum emission from both the
illuminated star and the disk peaks in the UV and the disk contributes a
greater fraction than at optical wavelengths (Cheng, Vrtilek, \&\ Raymond
1995).  From models of the variable UV continuum as observed with the
International Ultraviolet Explorer ({\it IUE}, Vrtilek \&\ Cheng 1996)
showed that a change in the accretion disk precession could explain an
anomalously low period of X-ray emission. 

To study global accretion one should observe gas
at ionization stages and temperatures resulting from X-ray illumination of
the disk, and one would need to resolve Doppler-shifted velocities that
correspond to the orbital motion of the neutron star (160 km~s$^{-1}$),
motion in the accretion disk (expected to be $\approx300$~km~s$^{-1}$ at
the edge of the disk), and the velocities expected in a disk or stellar
wind ($\sim1000$~km~s$^{-1}$). 

The strong UV resonance lines from N\,{\sc v}, Si\,{\sc iv}, and C\,{\sc
iv}, first seen with {\it IUE} (Howarth \&\ Wilson 1983b), presumably
result from photoionization of the accretion disk and HZ~Her (Raymond
1993, Ko \&\ Kallman 1994). These lines are much stronger than the
strongest optical high ionization line, He\,{\sc
ii}$\lambda4686$, both in absolute flux and equivalent width. 
Observations with the Faint Object Spectrograph (FOS) aboard the 
Hubble Space Telescope ({\it HST}) showed that these lines are still 
present at a few percent of maximum brightness during mid-eclipse when the 
disk and heated star should be entirely obscured (Anderson et al. 1994). 
The source of this emission may be an expanding wind.

The Goddard High Resolution Spectrograph (GHRS) on {\it HST} first
resolved these emission lines at $\approx35$~km~s$^{-1}$ resolution
(Boroson et al., 1996) to discern variable broad and narrow emission 
components. 

The {\it HST} Space Telescope Imaging Spectrograph (STIS) confirmed
that the resonance lines have at least two components (Vrtilek et al.
2001). A broad component arises on the accretion disk while a narrow line
component may be associated with HZ~Her. Prior to the STIS
observations, the accretion disk had only been observed as it contributed
to the continuum light curve or obscured the central X-rays. During
eclipse ingress and egress, the broad lines seen with STIS behaved
as expected for lines from an accretion disk rotating prograde
with the orbital direction.  The blue edge of the line was obscured
first in eclipse ingress and appeared first in egress. 

Observations with the Far Ultraviolet Spectroscopic Explorer ({\it FUSE})
are complementary to the existing {\it HST} observations. Long-term
variability in the system makes it impossible to combine 
rigorously 
data from different epochs.  Analyzed separately, however, the {\it FUSE}
wavelength range of 900-1200\AA\ offers similar advantages and powerful
consistency checks to analysis of the {\it HST} bandpass of 1200-1700\AA.
Both wavelength ranges have strong resonance lines that respond to X-ray
photoionization.  Observations with the Hopkins Ultraviolet Telescope
({\it HUT}) showed that the O\,{\sc vi} doublet has flux comparable to the
N\,{\sc v} doublet, the brightest near UV line (Boroson et al. 1997). 

The resonance line doublets offer optical depth information through their
doublet ratios, but if the lines are as broad as the doublet separation,
it may be impossible to determine the individual contribution of each line
where they overlap. The separations of C\,{\sc iv}$\lambda\lambda
1549,1551$, Si\,{\sc iv}$\lambda\lambda\lambda 1393,1403$, N\,{\sc
v}$\lambda\lambda 1238.1242$, O\,{\sc vi}$\lambda\lambda 1032,1038$,
S\,{\sc vi}$\lambda\lambda 933,944$ correspond to Doppler shifts of 500,
1900, 960, 1650, 3600 km~s$^{-1}$, respectively. The far UV lines O\,{\sc
vi} and S\,{\sc vi} compare favorably with the near UV lines (i.e. have
greater separation and are less likely to overlap), except for Si\,{\sc
iv}, which suffers from confusion with an O\,{\sc iv} blend near 1400\AA. 

For the present observations, we observed an entire 1.7~day binary orbit
with {\it FUSE}. A major goal of this program was to apply the method of
Doppler tomography (\S7), which has the advantage of
diagnosing the accretion flow of different systems without bringing to
bear more than a few assumptions. 

Table~1 shows the measured physical parameters of the system and
parameters we adopt for our models. 

\section{FUSE Observations}

With {\it FUSE} (the Far Ultraviolet Spectroscopic Explorer) we extend
observations in the UV spectral range to 900-1190\AA, a range observed
only once before using {\it HUT}, the Hopkins Ultraviolet Telescope aboard
the ASTRO-1, carried aboard the Space Shuttle but not placed in orbit
(Boroson et al. 1997). While {\it HUT} had a resolution of $\approx3$\AA,
{\it FUSE} has a resolution of $\approx0.05$\AA. 

{\it FUSE} is a NASA {\it Origins} mission operated by The Johns Hopkins
University. Four aligned telescopes feed two identical far-UV
spectrographs. With resolution R$=20000$ FUSE approaches {\it HST} in its
utility for our program; the time coverage of the {\it HST} Space
Telescope Imaging Spectrograph (STIS) was limited because the
detectors were turned off when the spacecraft passed through the South
Atlantic Anomaly. The FUSE mission is described in more detail in Moos et
al. (2000) and its on-orbit performance is described in Sahnow et.  al.
(2000). 

Our {\it FUSE} observations began on June 9, 2001 at 7:47~UT. Table~2
shows the log of exposures, each of which is integrated over each {\it
FUSE} orbit of the Earth, with gaps when Her~X-1 goes below the horizon.
We use the orbital ephemeris of Deeter et al. (1991) to determine the
orbital phases of our observation. The exposure times listed in Table~2
are in most cases equal to the raw observation time. However, for cases
such as observation 13, interrupted by a passage through the South
Atlantic Anomaly and not occultation by the Earth, the Exptime listed is
in general the minimum good exposure time for any {\it FUSE} detector. 
The data were obtained through the LWRS aperature and in TIMETAG mode.

We used the CalFUSE pipeline software version 3.0.8 to extract and
calibrate the data from all four {\it FUSE} telescopes.  Below 1100\AA, 
where emission features are sharp, we added an
offset to each wavelength scale, in intervals of 0.025\AA, so that the
absorption and emission features from each detector best agreed. 
We tested our wavelength calibration against the interstellar 
Si\,{\sc ii}$\lambda1020.6989$ absorption line, which we found to have a mean
heliocentric velocity of $-30$~km~s$^{-1}$. The standard deviation of 
the centroid of this line from orbit to orbit was $\approx0.03$\AA or 
$<10$~km~s$^{-1}$.

The S/N of the data was $\approx5$ per 0.1\AA\ pixel in the continuum in the
region $<1000$\AA\ and $\approx10$ near 1100\AA. The S/N within the 
O\,{\sc vi}
line had greater variation with orbital phase, as the doublet changed both
in shape and in strength. At $\phi=0.75$, the peak S/N within the doublet
was $\approx15$ per 0.1\AA\ pixel, while at $\phi=0.5$, the peak S/N was
$\approx25$ per pixel. 

For further analysis, we skip the region between 1120--1160\AA\ in the 1B 
LiF detector. This region suffers from a systematic decrease in counts 
known as the worm..

In Figures~1 and 2, we show average observed FUSE spectra at orbital
phases when the emission is dominated by the disk and star, respectively.
For the disk-dominated spectrum, we use observation number 20, at
$\phi=0.917$.  For the star-dominated spectra, we average observations
number 7 through 11 (phases $\phi=0.43--0.60$). 

For the {\it FUSE} spectra at orbital phases dominated by disk emission,
we compare with the STIS spectrum (observation root 
name O4V401010) during a Short-On state at $\phi=0.904$. 

For the spectrum dominated by star emission, we use the STIS
observation with root name O4V452050, observed during a Main-On state
approximately 5 months earlier, on January 24, 2001, starting at MJD
51933.779933 and lasting 2620 seconds. The mean orbital phase of this
exposure was $\phi=0.455$. This spectrum has not previously been
published. 

We compare the spectra with the time-weighted average of 
our continuum models (described in \S5). We
indicate prominent UV emission lines from the star system.

Geocoronal (airglow) lines produced in the Earth's atmosphere are also 
present (see Feldman et al., 2001): H\,{\sc i}\,Lyman-$\gamma$ 973, 
O\,{\sc i}\,5 989, H\,{\sc i}\,Lyman-$\beta$+O\,{\sc i}\,4
1026-1027, and N\,{\sc i} 1167.

\section{Interstellar Lines}

%
%

Interstellar features in the {\it FUSE} spectra are interesting not only 
for the direct information they provide on the interstellar medium (ISM), 
but a proper accounting of these features can remove systematic errors 
from the analysis of the line and continuum emission from the system itself.

The neutral Hydrogen column density, N$_H$, has previously been determined
to be $\approx10^{20}$~cm$^{-2}$ (Boroson et al. 2000) from the wings of
the saturated Lyman~$\alpha$ line as observed with the {\it HST} STIS. 
The {\it FUSE} bandpass includes further saturated lines in the Lyman
series, and these are consistent with N$_H\approx10^{20}$~cm$^{-2}$. 
This N$_H$ is also consistent with the E(B-V) value of 0.018 according to 
the Bohlin (1975) relation.

From the galactic latitude of 37.52$^\circ$ we should expect a sightline
with less H and H$_2$ than typically seen through the galactic disk. 
Indeed, while absorption lines from rotational levels $J=0$ through $J=3$
are readily identified, they are not saturated.  Thus the level
populations should be easy to measure and it should be easy to compensate
for the effects of the absorption lines on the spectra. 

In Figure~3, we show a patch of the time-averaged {\it FUSE} spectrum,
with the absorption profiles expected from source with a flat spectrum
absorbed by columns of (2.8, 7.5, 4.1, 3.4)$\times 10^{14}$ cm$^{-2}$ for
absorption from rotational levels $J=0,1,2,3$, respectively.  We generate
the H$_2$ profiles from the templates of McCandliss (2003), which are
based on Abgrall et al. (1993a,b).  We have assumed a line velocity
parameter $b=10$~km~s$^{-1}$ and have convolved the profiles with a
Gaussian to simulate the {\it FUSE} Line Spread Function. 

The two lowest rotational energy levels should have populations given by 
a Boltzmann factor, taking into account the statistical weights of the 
two levels, $g_1/g_0=9$.  The energy difference between the levels is
$\Delta E_{01}=170.5$~K.  We then estimate that the temperature of the 
intervening H$_2$ gas is $T_{01}=140$~K.  This is hotter than found for 
H$_2$ gas from disk stars, consistent with the high galactic latitude
of the line of sight (Gillmon, Shull, Tumlinson, \&\ Danforth 2006).

The average molecular fraction, defined by
\begin{equation}
f_{\rm H2}=\frac{2\mbox{N}(\mbox{H}_2)}
{
\mbox{N(H I)}
+
2\mbox{N(H2)}
}
\end{equation}
is therefore $\gae4\times10^{-5}$ from the observed H$_2$ lines, which 
have N(H$_2$)$\approx2\times10^{15}$~cm$^{-2}$.  
Ratios of higher rotational levels than $J_1/J_0$ are determined not by 
temperature and collisional excitation but by background FUV radiation. 
The H2 column density observed for Her~X-1 places it just above the 
boundary at which H2 clouds start to become optically thick to the FUV 
radiation.  Gillmon et al. (2006) state that $f_{\rm H2}\approx10^{-5}$ 
is a typical molecular fraction below this boundary.

Thus the first four levels of rotational excitation of H$_2$ appear to 
match observed absorption features and to be caused by H$_2$ clouds that 
are not out of the ordinary.

\section{Photometry of Emission Lines}

The lines vary with the binary orbit in a manner similar to that
previously observed in the near UV with {\it HST}. The flux peaks 
generally near $\phi=0.5$, when the X-ray heated face of HZ~Her
points toward the observer, although there may be a dip very close to
$\phi=0.5$ as the accretion disk occults the star.

In Figure~4 a through i respectively, we show the photometric variation
with orbital phase of the emission lines S\,{\sc vi}$\lambda 933.4$,
S\,{\sc vi}$\lambda944.5$, C\,{\sc iii}$\lambda 977$, N\,{\sc
iii}$\lambda991$, O\,{\sc vi}$\lambda1031.9$, O\,{\sc vi}$\lambda 1037.6$,
S\,{\sc iv}$\lambda 1073$, P\,{\sc v}$\lambda 1128$, and C\,{\sc
iii}$\lambda 1176$. We have subtracted the background flux using a model
that we present in \S5.  For the N\,{\sc iii} line, we remove the flux of
nearby airglow lines. 

In Table~3, we show the cross-correlation coefficients between the
measured lightcurves of the spectral lines and the lightcurves of the disk
and star contributions to the continuum lightcurve, as determined by our
continuum model. 

\section{UV Continuum Fits}

The UV continuum varies, as the optical continuum does, with the 1.7 day
orbital period.  The continuum generally peaks near $\phi=0.5$, when the 
X-ray heated face of the Roche lobe points towards the viewer, but a 
portion of the normal star may be blocked by the accretion disk, and 
depending on long-term phase, the actual peak may occur within 
$\phi\pm\Delta \phi\approx 0.5\pm 0.1$.  

The optical/UV orbital light curves do not repeat exactly over the 35-day 
disk precession cycle. The disk may block X-rays from heating portions of 
the star, and as it precesses it projects varying areas into the line of 
sight. The precessing disk should have different eclipse light curves 
throughout the 35 day cycle.

To fit the UV continuum observed with {\it FUSE}, we assume that X-ray
heating of HZ~Her and the accretion disk cause the entirety of continuum
emission.  The details of our simulation are similar to those of Vrtilek
et al. (1990, especially the appendix), and Howarth \&\ Wilson (1983a)
(which was an elaboration on an earlier analysis of Gerend \&\ Boynton,
1976), and binary simulation codes by Wilson \&\ Devinney (1971).  The
method is summarized in Appendix~A. 

We fit the spectrum using a reddening $E(B-V)=0.018$ and for the
extinction curve, we use Equation~5 in Cardelli, Clayton, \&\ Mathis
(1989). 

We also use a more modern determination of the distance to Her~X-1 
(Reynolds et al. 1997). This greater distance requires a larger $\dot{M}$ 
to reach the same continuum flux. Thus the scale of our $\dot{M}$ values 
are large compared with those reported earlier from {\it IUE} and {\it 
HST} observations. The {\it IUE} observations found a range of $-\log 
\dot{M}=8.30$ to $8.68$ (given changes in the long-term phase during an 
anomalous low state), whereas the {\it HST} found $-\log 
\dot{M}=8.19\pm0.06$.

When fit our continuum model to the observed {\it FUSE} spectrum from each 
{\it FUSE} observation, we find a range of $-\log \dot{M}=8.12$ to 8.38.

For our fits we ignore wavelength regions that contain prominent 
emission or absorption lines. We correct for absorption from the 
first 4 rotational energy levels of interstellar H$_2$ by multiplying
the spectrum by the model presented in \S3.

Table~3 gives the best-fit values of the mass accretion rate, $\dot{M}$, 
which we allow to vary as a free parameter for each {\it FUSE} orbit. We 
also list the reduced $\chi^2$.

We have used our continuum fits to examine the continuum photometry.  The
model allows us to separate the continuum emission
into contributions from the heated face of
HZ~Her and from the accretion disk. In Figure~5 we show the model flux 
near the O\,{\sc vi} lines split into disk and stellar components.


\section{Eclipse Models}

We made a simple model for emission from a symmetric precessing
accretion disk undergoing eclipse by the Roche lobe of its companion. We
then allowed free parameters of that model to vary in order to fit the
observed O\,{\sc vi} profiles during eclipse ingress and egress.

The disk and Roche lobe geometry are based on the model of Howarth \&\
Wilson (1983a), while the formation of the emission lines follows the model
of Horne (1995), and is described in Appendix~B. We allow as free parameters 
the exponent of a radial
power law of optical depth and elements of the Mach turbulence matrix. We
allow the normalization of the line flux at each orbital phase to vary as
a free parameter.

We calculate the eclipsing edge of the Roche lobe of 
HZ~Her assuming corotation with the orbit.

Before fitting the model to the data, we subtracted the mid-eclipse
spectrum from all spectra. The origin of the UV mid-eclipse spectra was
explored by Anderson et al. (1994) who concluded that it probably does
not arise in the accretion disk. The mid-eclipse O\,{\sc vi} lines are 
broad, extending from -200 to $+500$~km~s$^{-1}$ heliocentric velocity.

We also apply narrow gaussian absorption lines to our model spectra in
order to simulate a possible C\,{\sc ii} interstellar line near 1036\AA, a
possible C\,{\sc ii}$*$ line near 1037\AA, and a O\,{\sc i} line near
1039\AA. We also include narrow interstellar absorption near the rest
wavelengths of the O\,{\sc vi} doublet. A narrow absorption line can be
seen near the blue component of the doublet at $\lambda=1031.9$\AA, but
the weaker line at 1037.6\AA\ cannot be seen explicitly. We fix the
wavelengths, widths, and optical depths of these lines and do not let them
vary in our fit. 

We fix all the parameters describing the disk and orbit to those given in 
Howarth \&\ Wilson (1983a), except the outer radius we fix to 
$2\times10^{11}$~cm, following Cheng et al. (1995).

The results are shown in Figure~6.

The reduced $\chi^2$ of the fit was 2.8 with 1418 degrees of freedom.

The power law index for the radial dependence on optical depth was 
$\alpha=-1.2$, between the $\alpha=-1.5$ result found for the N\,{\sc v} 
doublet by similar methods in Boroson et al. (2000) and 
the $\approx-0.55$ value expected from 
the simulations of Raymond (1993).

Although we based our fits on the model of Horne (1995) which allows for
anisotropic turbulence, the fits do not unambiguously settle on
particular values of the Mach matrix. Here, for simplicity, we present a 
model with isotropic Mach=1 turbulence.

As with the resonance doublets in the near UV, the observed
blue component of O\,{\sc vi} near eclipse is stronger than the red 
component, 
which suggests that much of the line flux is formed where $\tau_0\lae1$. 
However, the doublet ratio is uncertain because strong interstellar 
C\,{\sc ii} 
and O\,{\sc i} absorption lines affect only the red O\,{\sc vi} doublet 
component. 
Weaker interstellar lines, which we have not modeled, may be present as 
well, and may preferentially absorb one or the other O\,{\sc vi} component.

The model lines are double-peaked outside of eclipse, but one peak of the
red component is absorbed by interstellar O\,{\sc i} absorption. The observed
line profiles are double-peaked at $\phi=0.84$ and double-peaked but 
broader at $\phi=0.80$. Observations at $\phi=0.35$ and $\phi=0.39$, when 
narrow line emission predominates, show a dip in the flux at the position 
of the gap between peaks from the accretion disk line.

The line normalizations we found from our disk eclipse model can be
compared with the mass accretion rates inferred from our continuum fits.
They do not appear correlated.  At some of the phases investigated, we
observe only a small fraction of the accretion disk, and none of the
illuminated star, so that systematic errors in the line or continuum
models may make accurate estimates of $\dot{M}$ impossible. In particular,
we have considered the disk flat when it is probably warped, we have not
calculated the ionization structure of the disk, we have ignored possible
emission by the accretion stream, and we have ignored the role of winds in
this system. 

Chiang (2001) presented models of accretion disk eclipse in Her~X-1 that
considered a disk wind. These models were motivated by the lack of
double-peaked line profiles in any observations of UV resonance lines.
Although the current model predicts double-peaked emission lines when the
full disk is visible, the S/N is not strong 
enough to rule out a double-peaked structure. In addition, interstellar 
absorption lines coincide with three of the four peaks of the O\,{\sc vi} 
doublet. 

It is clear that a model of a partially eclipsed
accretion disk with parameters considered standard from previous work
provides an excellent match to this data set.  There appear to be 
too many ways to improve the current fit to justify singling one out. 

\subsection{Anomalous emission at $\phi=0.876$}

The fit at $\phi=0.876$ is particularly poor and we suggest that
a process in addition to the partial eclipse of a Keplerian accretion 
disk affects the line profile.

Figure~6 shows that even though more of the disk is eclipsed than at 
$\phi=0.835$, the emission lines at $\phi=0.876$ are actually stronger. 
There also appears to be an additional absorption component near the rest 
wavelength of O\,{\sc vi}$\lambda1032$.

The continuum at $\phi=0.876$ has not increased over the continuum at
$\phi=0.835$ as much as the line emission has increased.  This is
consistent with the earlier result that the X-ray illumination from our
line and continuum models at other phases were not correlated. 


\section{Doppler Tomography}

The Doppler tomography method, developed by Marsh and Horne (Marsh \&\
Horne, 1988, Marsh 2005) takes as input an emission line which is
broadened by line of sight motion. The line must be observed over a good
sampling of the binary orbit. In analogy with medical tomography, a
successful Doppler tomogram builds a higher-dimensional view of the object
from different observational slices. 

Analysis of 
Doppler shifted emission lines is prone to the confusion of different 
gas sources which have been projected onto the same observed velocity. 
The tomographic method presents a transform of the same data into
features separated not only by projected velocity, but also by orbital 
variability.

The output tomogram is not a physical image of an accretion disk, but 
rather an image in ``projected velocity space.'' A component of the 
emission that varies sinusoidally in velocity (but remains constant in 
flux) is placed at a point in velocity space 
along a circle with a radius equal to the orbital velocity of that 
emission component. The phase of the sinusoid determines its position 
along that circle.

The method assumes that all of the variation in a spectral line must be 
caused by the
orbit changing the projection of the motion into the line of sight. (For
example, the change in a spectrum must not be caused by a bright spot
being obscured by an opaque object.) Even outside of the conditions in
which it strictly applies, however, tomographic images can provide a
common view of the data and may suggest directions for further analysis. 

Technically, a Doppler tomogram is an inversion of

\begin{equation}
\label{eq:tomogram}
f(v,\phi)=\int_{-\infty}^{\infty} \int_{-\infty}^{\infty} I(v_x,v_y) 
g(v-v_R) d\,v_x d\,v_y
\end{equation}

which gives the flux $F(v,\phi)$ at the projected velocity $v$ and the 
orbital phase $\phi$ as a summation of the velocity space tomogram 
$I(v_x,v_y)$ with a line broadening function $g(V-V_R)$ which is usually 
narrow. This integral is an instance of a Radon transform. The line 
broadening function for our FUSE spectra 
is approximately 0.05\AA and for tomographic analysis we assume it is a 
delta function.

The relation between the radial velocity and the two axes of the Doppler 
tomogram, $v_x$ and $v_y$ is given by

\begin{equation}
\label{eq:doppler}
v_R=\gamma - v_x \cos 2 \pi \phi + v_y \sin 2 \pi \phi
\end{equation}
where $\gamma$ is the systemic velocity of the system.

The formation of the spectrum from the tomogram can be thought of as 
follows. The phase $\phi$ determines a direction in the velocity space 
plane, and moving along that line, summing the intensity perpendicular 
to the line, one ideally forms the spectral line. At $\phi=0.25$, one 
views from the right in our figures, at $\phi=0.5$ one views from the 
bottom, and at $\phi=0.75$ one views from the left.

In this paper, we determine $I(v_x,v_y)$ from $F(v,\phi)$ by means of
Fourier-Filtered Back Projection, which inverts 
Equation~\ref{eq:tomogram} through
\begin{equation}
I(v_x,v_y)=\int_{0}^{1} \bar{f}(\gamma-v_x \cos 2 \pi \phi+v_y \sin 2 \pi 
\phi,\phi)  d\,\phi \end{equation}
where to calculate $\bar{f}(v,\phi)$ from $f(v,\phi)$, one takes its 
Fourier transform, multiplies by a ramp filter and a Wiener (optimal) 
filter based on the noise level, and then takes the inverse Fourier 
transform.

Although it may seem as if this filtering complicates the method of
back-projection, a ramp filter (multiplying the Fourier transform by the
frequency) is required in order for back projection to produce a rigorous 
inversion of Equation~\ref{eq:tomogram}.
Otherwise, back projection would only
produce a smeared version of the true tomogram. 

Unfortunately, using a ramp filter also magnifies pixel to pixel noise. We
apply a standard Wiener filter that assumes a signal amid a noise
component that is constant with frequency. We find the power spectrum of
the line spectra is well fit by several components of the form
$P(\omega)=P_0 \exp(-k \omega)$ (with $k>0$).  The contribution of the
signal is negligible at high frequencies $\omega$, where the power
spectrum is fit well by a constant noise level $n$. The action of the
Wiener filter is then to multiply the power spectrum by
$P(\omega)/(n+P(\omega)$, which is nearly 1 when the signal dominates the
noise and negligible at high frequencies where the spectrum is almost
entirely noise. We used the same form of $P(\omega)$ for each line at each
orbital phase, and made a visual comparison between the actual power
spectrum and the function $P(\omega)$.  Because the filter adapts to
different noise levels, the amount of smoothing varies from line to line. 
For the lines with highest S/N, such as O\,{\sc vi}$\lambda1032$, the
turnover frequency for the smoothing filter is on the order of 3 pixels,
whereas for weaker lines such as S\,{\sc vi}$\lambda 944$, the smoothing
is on the order of 10 pixels. 

We did not apply the popular method of Maximum Entropy regularization,
which awards higher probabilities to those tomograms that provide the
least information in the sense that they are smoothest. 

\subsection{Applications of Doppler Tomography}

In Figures~7--14 we show Doppler tomograms of the emission lines of
S\,{\sc vi} at 933 and 944\AA, N\,{\sc iii}$\lambda 911$, O\,{\sc vi} at
1031\AA\  and 1037\AA, P\,{\sc v}$\lambda 1128$, S\,{\sc iv}$\lambda1073$,
and C\,{\sc iii}$\lambda1175$.  The tomogram of C\,{\sc iii}$\lambda977$
is not shown as it is contaminated by the presence of saturated
absorption. We have interpolated between orbital phases for the
integration, which we perform between $\phi=0.2$ and $\phi=0.85$. The
grayscale of the tomograms extends linearly from the minimum to the
maximum. 

To understand the C\,{\sc iii} multiplet, we consult {\it 
CHIANTI}, a software package and atomic database described in
Dere et al. 1997 and Landi et al. 2006. The database, 
used extensively by stellar and solar astrophysicists, contains energy 
levels, wavelengths, radiative transition probabilities, and excitation 
data for many ions. The associated software package, written in IDL 
(Interactive Data Language), can be used to examine how line ratios vary 
with temperature and density, subject to certain limiting assumptions.
We find that C\,{\sc iii} contains 
emission components at $0, 85, 168, 199, 269$, and 367 km~s${-1}$ 
relative to the 1175.59\AA\ component. 

We were able to reduce the artifacts by deconvolving with a
triple-gaussian $g(V-V_R)$ in Equation~\ref{eq:tomogram}. We tried
Gaussians of equal weight at velocities $(0,+168,+367)$ km~s$^{-1}$
(Figure~14) and find there is a single bright spot near the Roche lobe. 

A further problem with the C\,{\sc iii} multiplet is that the model of the
continuum in this region predicts an absorption dip. To reduce artifacts,
we use the simpler and smoother Kurucz model atmospheres, with 1\AA\
resolution for all of the tomograms, instead of the models based on actual
stellar spectra observed with {\it FUSE}, which we have computed with
0.1\AA\ resolution and which include some counting statistics noise.

The signal in the tomograms, except as noted above, is concentrated in a
peak offset from the Roche lobe of HZ~Her. The traditional signature of an
accretion disk is not obviously present. Accretion disks are expected to
cause broad line emission that appears as a ring in a tomogram, with
diminished flux within the velocity at the edge of the accretion disk, or
about 300~km~s$^{-1}$ for Hercules~X-1.  If the disk emission is
symmetric, we expect it to be centered on the position of the neutron star
in velocity space, $(vx,vy)=(0,-169)$. 

\subsection{Narrow Emission and Broad Absorption Components}

There is a simple interpretation of the tomograms in terms of P~Cygni
lines commonly seen in stellar winds from massive stars (and in {\it HST}
spectra of Her~X-1, Boroson, Kallman, \&\ Vrtilek 2001). P~Cygni lines, 
caused by resonance scattering, are characterized by red-shifted emission 
and absorption blue-shifted by velocities common in the 
wind. There is a hint of blue-shifted absorption in the trailed 
spectrograms at $\phi=0.3-0.6$, although there may be an interstellar 
absorption line as well at these wavelengths and the gap between the 
peaks of the accretion disk spectrum could also appear to be absorption.

We note that for many tomograms in addition to the bright spot at
$(vx,vy)\approx(150,-30)$, there is a dark spot at
$(vx,vy)\approx(-100,50)$.  (For O\,{\sc vi}$\lambda1031$, this absorption
may have a darker lobe near the rest wavelength from interstellar O\,{\sc
vi} absorption.) Near the start of the phase range used by our tomograms,
at $\phi\approx0.3$, the observer's line of sight comes from the lower
right in Figure~7. As the line of sight passes through both bright and
dark spots, the observer at $\phi\approx0.3$ sees a diminished central
peak in the spectral lines. 

At $\phi\approx0.5-0.6$, the viewer looks at the system from the bottom
of the plot, stacking pixels vertically. The result is red-shifted
emission (the bright spot) and blue-shifted absorption (the dark spot). 

Although the blue-shifted absorption at $\phi=0.3-0.6$ may be explained 
through the analogy of P~Cygni lines, the red-shifted emission at these 
phases does not fit this explanation. In a P~Cygni line, except when a 
very hot wind is present (as in a Wolf-Rayet star), all the emission 
is caused by spherically symmetric scattering, so that 
conservation of photons requires that the forward-scattered flux be 
nearly equal to the back-scattered (``absorbed'') flux (some portion of 
the forward-scattered emission is blocked by the star itself). 
In contrast, any blue-shifted absorption in the Her~X-1 spectrum is only 
a small fraction of the emission. Further, P~Cygni emission is only 
red-shifted because it overlaps with the absorption on the blue end; the 
actual emission spectrum should be symmetric about zero velocity.


It seems most natural to associate the bright spot with the X-ray heated 
face of the normal star, as it follows roughly the same light curve and 
appears near the star and far from the disk on the tomograms. Yet 
the placement of the spot, or equivalently, the orbital variation with 
velocity, cannot be accounted for easily by such a model.

In Figure~15, we show the observed spectrum near O\,{\sc vi} (bold), the
spectrum expected if the illuminated surface of HZ~Her emits O\,{\sc vi}
Doppler shifted by the local rotational velocity (dashed), and an
empirical Gaussian model of the emission lines. The Doppler shifts
predicted from the rotation of the surface of HZ~Her are less than
observed in the narrow emission lines. 

To model the narrow O\,{\sc vi} emission lines we adapt our model of the
continuum emission. This model takes into account the X-ray shadow cast by
the disk and the eclipse of portions of the normal star by the disk.  The
reprocessed O\,{\sc vi} emission is proportional to the incident X-ray 
flux. We
also experimented with a model in which the X-ray illumination was
non-isotropic, illuminating HZ~Her with greater luminosity at $\phi=0.25$
than at $\phi=0.75$. This model still does not account for the observed
velocities. 

We also made an empirical model of the lines as Gaussians. The flux in the
Gaussian line emission at each phase was proportional to the flux in the
stellar component of our continuum model using Kurucz model spectra. The 
central velocity in this model varies
sinusoidally with phase. We added Gaussian absorption to correspond to
features seen in {\it HST} STIS spectra of the N\,{\sc v}$\lambda1240$,
Si\,{\sc iv}$\lambda1400$, and C\,{\sc iv}$\lambda1550$ doublets near 
$\phi=0.3-0.6$.
We set the maximum projected redshift of the emission lines and blueshift
of the absorption to occur near $\phi=0.5$, but delayed by $\phi=0.06$,
corresponding to a deflection of a wind by the Coriolis effect. The
absorption line has maximum covering fraction near $\phi=0.5$ that
smoothly varies to 0 at $\phi=0$. (Any wind is probably confined to the 
cylinder from the star towards the disk, as the emission lines seen in 
mid-eclipse are only $\sim1$\%\ of the peak line strength, as observed 
with the {\it HST} FOS by Anderson et al. 1994). Both emission and 
absorption components
have constant width. Both emission and absorption components vary about a
central velocity given by the L1 point. 

When these features are added to the model of the disk spectrum in 
\S6, assumed to move with the neutron star's 
169~km~s$^{-1}$ orbit, and random counting statistics noise is added, the 
resulting tomogram (Figure~16) resembles the observed tomogram, while 
showing only a hint of the presence of an accretion disk. Without the 
assumed phase delay of $\Delta \phi=0.06$, the bright spot would appear 
rotated to $vy=0$.

\section{Line Ratio Diagnostics}

The {\it FUSE} spectra show emission lines that may serve as diagnostics
of density, temperature, or optical depth.

\subsection{C\,{\sc iii} line diagnostics}

The simultaneous measurement of C\,{\sc iii}$\lambda977$ and
C\,{\sc iii}$\lambda1176$ can provide us with an important diagnostic. 
The C\,{\sc iii} 1176/977 ratio is sensitive to
density in the range up to about $n_e=10^{11}$~cm$^{-3}$. At higher
densities (up to $n_e\sim10^{15}$), the populations of the ground and
metastable levels are entirely determined by collisions.  In that
regime, the 1176/977 ratio depends only on temperature and the optical
depths in the lines.

We ignored flux from the {\it FUSE} LiF2B detector near the C\,{\sc
iii}$\lambda977$ line, even though this detector recorded flux down to
exactly 977\AA. The edge of the detector causes the flux measured in this
region was significantly lower than the other two detectors, so that
including this data would have created an artificial dip at wavelengths
$\lambda>977$\AA. 

The fluxes we measure for C\,{\sc iii}$\lambda977$ are lower limits
because of a saturated interstellar absorption line at the rest
wavelength.  We correct the fluxes for interstellar reddening using
E(B-V)=0.018. 

We compare the fluxes of the two lines near $\phi=0.6$, when the fluxes
peak. If the C\,{\sc iii}$\lambda977$ line behaves similarly to the
O\,{\sc vi} lines, then the bright narrow emission will, near this phase,
have the greatest redshift. If so, the line may be redshifted away from
the interstellar absorption and the flux measurement may be more accurate
near this phase. If the O\,{\sc vi}$\lambda1032$ line were subject to
similar interstellar absorption lines as the C\,{\sc iii}$\lambda977$
line, the flux at $\phi=0.6$ would be diminished by $\sim$20\%. 

If we measure the raw C\,{\sc iii}$\lambda977$ flux from the two phases
surrounding $\phi=0.6$ to be $7\pm1\times10^{-13}$~erg~s$^{-1}$~cm$^{-2}$,
then we find a 1175/977 ratio of $0.82\pm0.16$. If we measure the C\,{\sc
iii}$\lambda977$ flux at exactly the peak at $\phi=0.6$ and attempt to
correct for interstellar absorption lines, we find a ratio of
0.60$\pm0.09$. 

These ratios are at the high end of those predicted by the {\it CHIANTI}
software, which uses atomic data from the {\it CHIANTI} database and
computes line ratios, given certain simplifying assumptions (for example,
the lines are optically thin and the gas is collisionally ionized and not
photoionized). If we consider a density of $n_e=10^{13.4}$~cm$^{-3}$, the
lower ratio found above would restrict the temperature to
$T>6\times10^4$~K. 

The ratio is rendered uncertain observationally by the saturated 
interstellar absorption line, and theoretically
because of optical depth effects. Raymond (1993) presented models of line
emission from an X-ray illuminated accretion disk which took into account
optical depth through an escape probability formalism. One version of the
model, ``COS", assumed cosmic abundances while the other, ``CNO", assumed
that abundances had been altered by CNO processing. The CNO models should
be more appropriate for Her~X-1.  In the two cases the $I(1176)/I(977)$
ratios were 1.3 and 1.1, respectively. The observed emission at 
$\phi=0.6$ may, however, have an origin not on the disk but on the 
normal star. 

\subsection{Bowen fluorescence and N\,{\sc iii}$\lambda991$ emission}

The Bowen Fluorescence process (Schachter et al., 1989) arises because of
the nearly perfect coincidence of the He\,{\sc ii} Ly$\alpha$ and the
O\,{\sc iii} 2p$^2$ -- 2p3d resonance line ($\lambda$304), 
resulting in
O\,{\sc iii} near--UV primary cascades at $\lambda\lambda$3133, 3444 and
secondary cascades at $\lambda$374 back to the ground state.  If
conditions are right, an additional fluorescence occurs, since the
\ion{O}{3}$\lambda$374 line is almost coincident with the two \ion{N}{3}
2p -- 3d resonance lines, resulting in N III optical primary cascades at
$\lambda$4634, 4641, 4642.  The emission in all of these cascades is
completely dominated by the Bowen process; detection of any of them is a
clear confirmation.  We note that in Her~X-1 the process requires
substantial optical depth ($\sim10^6$) in the \ion{He}{2} Ly$\alpha$ and
\ion{O}{3} $\lambda$374 {\em pumping} lines, because otherwise they will
simply escape without conversion to a Bowen line. 

In the Sun, Raymond (1978) discovered \ion{O}{3}$\lambda$304 Bowen
emission.  It was found that measurements of \ion{O}{3}$\lambda$703, a
{\em ground-state} Bowen cascade which, therefore, can also be produced
by collisional excitation, can serve as a density diagnostic, and compared
favorably with other solar estimates of $n_e$.  

We may use N\,{\sc iii}$\lambda$990, an analogous ground--state Bowen 
cascade produced as a result of the \ion{N}{3} primary cascades
(``$\lambda$4640''; all other \ion{O}{3} and \ion{N}{3} ground-state
cascades lie below the Lyman limit and hence are unobservable).


The intensity of N\,{\sc iii}$\lambda990$ resulting from Bowen 
fluorescence is related to the intensity of the Bowen lines at 4640\AA by
\begin{equation}
I_f(991)=(4640/991) B(\mbox{3p2p}) I(4640)
\end{equation}
where $B(\mbox{3p2p})$ is the branching ratio of 2p-2s$^2$p$^2$ 2D versus
3p-3s. From CHIANTI, we take $B(\mbox{3p2p})=0.55$ and 
$B(\mbox{3d3p})=0.0058$.

The remainder of the line intensity is therefore from collisional 
excitation and is given by $I_c(991)=I(991)-I_f(991)$.

Taking into account the reddening toward Her~X-1 implied by E(B-V)=0.018,
we find a peak flux in the 991\AA\ line of $8.6\pm0.7\times10^{-13}$
erg~s$^{-1}$~cm$^{-2}$. Non-simultaneous measurements of the optical
$\lambda$4640 line have a peak flux at similar orbital phases of
$7.0\pm0.2 \times 10^{-14}$ erg~s$^{-1}$~cm$^{-2}$ (Still et al. 1997).
The optical flux varies from orbit to orbit, and, moreover, probably
includes contributions from C\,{\sc iii} as well as N\,{\sc iii}.

These values imply that 20\% of the N\,{\sc iii}$\lambda991$ line flux is 
the result of the Bowen fluorescence mechanism.

We attempt to relate the density to the Bowen flux by
\begin{eqnarray}
\label{eq:bowen}
\frac{I_c(991)}{I_f(991)}  & = & 4\\
\nonumber & = & 
 \frac{n_e N(\mbox{N\,{\sc iii}}) q}{B(\mbox{3d3p}) B(\mbox{3p2p})
N(\mbox{N\,{\sc iii}}) \sigma I(374)}\\
\nonumber & = & 4.9\times10^8 n_e / I(374)
\end{eqnarray}
with $N(\mbox{N\,{\sc iii}})$ the density of N\,{\sc iii} and 
$q=6\times10^{-8}$~s$^{-1}$ 
the excitation rate of the 991\AA\ transition at T=30,000~K.

We estimate an upper bound on the O\,{\sc iii}$\lambda374$ intensity (the 
line
may be optically thick) from the O\,{\sc iii}$\lambda3132$ line observed with
the Faint Object Spectrograph on {\it HST} by Anderson et al. (1994). We
account for the other O\,{\sc iii} branches leading to the $\lambda374$ 
line by
using the observed O\,{\sc iii} Bowen spectrum of RR~Tel (Selvelli, Danziger,
\&\ Bonifacio 2007). We then estimate $I(374)\lae 0.032 (D/R)^2$, with
$D=6.6$~kpc the distance to the system and $R$ the radius of the emitting
region. Boroson et al. (2000) found 
$\log n_e=13.4\pm0.2$ 
for the narrow lines from the ratios of lines observed with {\it HST},
and Howarth \&\ Wilson (1983b) found $\log n_e=13.3$ from {\it IUE} 
observations.
Combining with Equation~\ref{eq:bowen}, we find $R\lae 7 \times
10^{10}$~cm. 

This is smaller than the size of the accretion disk 
($R_{\rm outer}\approx2\times10^{11}$~cm). We note that the tomograms show 
emission from a small region in velocity space.


Another test is provided by a comparison with theoretical models of
N\,{\sc iii}$\lambda991$ emission in the absence of the Bowen process. 
Raymond
(1993) found ratios between the N\,{\sc iii} line and C\,{\sc 
iii}$\lambda977$ 
for the COS model of $I(991)/I(977)=0.34$ and for the CNO model of
$I(991)/I(977)=0.53$. The observed ratio of $0.75\pm0.10$
requires enhancement by Bowen fluorescence and provides an excellent 
match to the CNO model if only 70\%\ of the N\,{\sc iii}$\lambda991$ line 
results from collisional excitation.

\section{Discussion}

The far UV spectrum of Hercules~X-1 shows clear evidence for a Keplerian 
accretion disk. A simple model such a disk can fit the broad O\,{\sc vi} 
emission lines near eclipse. The origin of brighter, narrow emission 
lines is still unknown. Doppler tomograms place this emission apart from 
the Roche lobe of HZ~Her.  

The bright narrow emission component generally follows the flux expected 
from the illuminated
portion of the normal star. However, it appears brighter than expected at
$\phi=0.2$ and the velocity excursion is also greater than is accounted
for by the rotation of the Roche lobe. A dense portion of a stellar wind
that moves along the surface of the star, or that originates at the L1
point and flows back towards the star, would be red-shifted at $\phi=0.5$
and blue-shifted at $\phi=0.25$ and $\phi=0.75$. Although such a model can
be made to reproduce the gross features of the line variation, there
remain too many possible alternate explanations for this to be convincing.
There could be a source of variable emission, absorption, or
scattering in addition to the normal star, for example, the gas stream or
the surface of the disk. 

In the presence of stellar winds, possibly transient or restricted in
solid angle, tomography did not show a clear signature of an accretion
disk in Hercules X-1. 

Although we found that a simple model of an accretion disk generally fit the
spectra during eclipse ingress and egress, there was an anomalous
observation in which the O\,{\sc vi} lines brightened as the disk eclipse
progressed. The phase $\phi=0.876$ is in the $\phi=0.8-0.9$ range in which
historically ``pre-eclipse dips'' of X-rays have been observed, and it is
tempting to relate the temporary brightening of the lines to a physical
cause or result of the dips. 

Alternately, an anomalous change to the emission lines could result from 
a change in shadowing of the X-ray emission. Boroson et al. (2001) 
concluded that anomalous {\it absorption} within the 
N\,{\sc v}$\lambda\lambda 1238,1242$ doublet changed too rapidly to result 
from material passing over the line of sight. The absorption changed so
rapidly probably because of a rapid change in  
the shadowing of the X-rays which ionize the gas.

Future X-ray spectroscopy missions such as {\it Constellation X} may be
able to resolve lines from more highly ionized species than the UV and Far
UV resonance lines (Vrtilek et al. 2004).
If more highly ionized
species track disk material better and are not as prevalent in the narrow
line region, we may be able to make reliable tomograms of the disk in this
or similar systems. Hydrodynamic models of this system may be required to
cut down on the vast parameter space of possible gas flows responsible for
the spectral signatures observed here. 

\acknowledgements

We would like to thank Jeff Bryant for helping us develop some {\it 
Mathematica} routines for Doppler tomography.  CHIANTI is a collaborative 
project involving the NRL (USA), RAL (UK), MSSL (UK), the Universities of 
Florence (Italy) and Cambridge (UK), and George Mason University (USA).
We would like to thank the anonymous referee for suggestions that 
improved the text and presentation, for pointing us to the S\,{\sc 
vi}$\lambda\lambda 933,944$ lines, and prompting us to analyze the 
interstellar molecular absorption lines.

\appendix

\section{Details of the Model of the Her X-1 Continuum}

With one free parameter, the mass accretion rate $\dot{M}$, we model 
the continuum emission from the accretion disk and X-ray illuminated face 
of the normal star. We do this not by means of radiative transfer 
calculations, but by calculating the heated temperatures of both surfaces, 
and co-adding spectra appropriate for those temperatures, either 
blackbody spectra, models of stellar spectra, or actual spectra of hot 
stars observed with {\it FUSE}.

Our model of the far UV continuum emission of Her~X-1 follows closely the 
methods of Vrtilek et al. (1990) and Cheng, Vrtilek, \&\ Raymond (1995).  
We describe the method here in detail, including the minor departures we 
have made to extend the spectral simulation to the far UV. 

The model of the accretion disk temperature as a function of radius, $T(r)$, 
includes the effects of both internal heating due to viscous forces and 
heating from X-rays from the neutron star.  

When we calculate the disk temperature as a function of radius, $T(r)$, 
we include both the heat  from viscous forces and X-rays from the 
neutron star.  The local energy generated by accretion is given by
\begin{equation}
\sigma T_0(r)^4 = \frac{3 G M \dot{M}}{8 \pi r^3}\left( 
1-\sqrt{\frac{R}{r}}\right)
\end{equation}
for Stefan-Boltzmann constant $\sigma$, neutron star mass $M$ and radius 
$R$, and disk mass accretion rate $\dot{M}$ (Shakura \&\ Sunyaev 1973).

For local energy balance, the energy emitted must equal the energy 
generated plus the X-ray energy absorbed:
\begin{equation}
\sigma T_{\rm eff}^4(r) = \sigma T_0(r) + \frac{a L_x}{4 \pi r}
\frac{\partial (h/r)}{\partial r}
\end{equation}
where we choose the albedo $a=0.5$, and the disk height at radius $r$ is 
$h(r)$.  The X-ray luminosity is related to the gravitational potential 
released by accreting matter:
\begin{equation}
L_{\rm x}=0.5 \frac{G M \dot{M}}{R}
\end{equation}

These equations, together with vertical hydrostatic equilibrium, can be 
solved numerically to determine $T(r)$. The surface of the noncollapsed 
star is also assumed to have an albedo $a=0.5$.

At each small region in the accretion disk, the temperature $T(r)$ is
compared with the critical disk temperature, $T_{dc}$. If $T(r)>T_{dc}$,
then we assume that the disk region radiates like a blackbody. If
$T(r)<T_{dc}$, then we interpolate through our library of actual stellar
spectra to find a spectrum appropriate for temperature $T(r)$.  From
experience fitting {\it HST} FOS spectra and fitting the Balmer jump, we
fix $T_{dc}=10000$~K so that in practice the disk almost always has
blackbody spectra, except at its edge where it is not illuminated by
X-rays. 

The stellar library from 1150 to 7500\AA\ is described in Cannizzo \&\ Kenyon
(1987).  The near UV spectra were obtained with {\it IUE} and are
described in Wu et al. (1982). All the spectra are normalized to have
constant flux in the V bandpass and are then scaled to their absolute
magnitudes using their V-R colors and the Barnes-Evans relation (Barnes,
Evans, \&\ Moffett 1978). 

For this paper, we have extended the library of stellar spectra further
into the UV by using {\it FUSE} spectra of main sequence stars between
O\,9.5 and A\,7.  These observations are described briefly in
Table~5.  Stars of temperature greater than $T_{\rm 
sc}=18,900$~K 
are not used in the fits to Her~X-1 and are not included in the table. We 
have interpolated over those wavelengths we
expect are affected by interstellar atomic or molecular absorption, and we
have applied a filter designed to smooth interstellar lines while
retaining stellar features.  We de-reddened each spectrum and scaled the
flux according to the star's visual magnitude V.  We have tested the
method and the use of these stellar spectra by using model stellar spectra
as predicted by Kurucz (1979). Plots of far UV spectra in our library, 
normalized to have a constant V magnitude, are shown in Figure~17.

For the purpose of computing the disk shadow, the disk is assumed to have a 
fixed opening angle and a fixed tilt 
from the orbital inclination, although the direction of the disk normal
precesses with the X-ray cycle in a manner described by Gerend \&\ Boynton
(1976) and Howarth \&\ Wilson (1983a).  Because of the finite disk 
opening angle, the outer regions of the disk can
occult the inner regions, and the disk edge itself can radiate, although
it is not illuminated by the central X-rays. 

The noncollapsed star is assumed to fill its Roche lobe, and its shape,
visibility, and illumination by the X-ray source are treated according to
methods evolved from Wilson \&\ Devinney (1971) to Howarth \&\ Wilson
(1983a). Following those references, we model the eclipse of the disk using
the Roche potential. We take into account both limb darkening and the
gravity-darkening appropriate for a late-type star. If the
temperature of some region on the surface of the star, after X-ray
illumination, is greater than some critical temperature $T_{sc}$, we
assume the region emits as a blackbody. If the local temperature is less
than $T_{sc}$, we again interpolate through our stellar library to find
the appropriate stellar spectrum with which the spot radiates.  From
earlier experience fitting Her~X-1 UV spectra, we fix $T_{sc}=18900$ (Cheng,
Vrtilek, \&\ Raymond 1995). 

The model, as currently realized, does not account for emission or 
shadowing by a gas stream between the noncollapsed star and disk.

\section{Details of the Model of the Disk Lines in Eclipse}

Accretion disks in theory have emission lines with a double-peaked shape,
with peaks separated by a velocity given approximately by the projection
into the line of sight of the orbital velocity at the edge of the
accretion disk. For Hercules~X-1, with typical neutron star mass, an 
nearly edge-on inclination, and an outer disk radius of $R\approx 
2\times10^{11}$~cm, we expect the peaks to occur at $\pm 
300$~km~s$^{-1}$. While greater velocities are found within the outer disk
radius, each ring at a fixed radius emits down to a Doppler velocity of 0
where the gas moves tangential to the line of sight. Regions of high 
Doppler velocity arise from progressively smaller regions in the disk, 
and contribute less to the overall line shape.

We have calculated detailed models of emission from an accretion disk and
have fit these to the profiles observed in Her~X-1 during eclipse ingress
and egress. In reality, the disk is likely to be warped, and in future
work we will test individual models of this warped shape. The importance
of the demonstration in this paper is that it shows how easy it is to get
general agreement with the observed line profiles given the standard
picture of the accretion disk. 

The method in detail follows that of Horne (1995), which allows for 
nonisotropic turbulence and for Keplerian shear, that is, the local 
dispersion in disk velocity as a result of the Keplerian flow itself. This 
introduces free parameters for the ``Mach matrix" describing the 
turbulent flow, in addition to the power law exponent that we use to 
as a phenomenological description the strength of the emission line as a 
function of radius in the disk. For simplicity, we keep the elements of 
the Mach matrix constant with radius in the disk.

Following Horne (1995), the local line profile is given by
\begin{equation}
I_\nu \propto (1-e^{-\tau_\nu})
\end{equation}
where the optical depth $\tau_\nu$ at frequency $\nu$ is given by
\begin{equation}
\tau_\nu=\tau_0 e^{(V-V_0)^2/2\Delta V^2}
\end{equation}
for a given line center optical depth $\tau_0$.

The nonisotropic turbulence and shear enter through $\Delta V$, given by
\begin{equation}
\Delta V^2  =  \Delta V^2_{\rm therm} + \Delta V^2_{\rm shear}
	+ \Delta V^2_{\rm turb}
\end{equation}
or
\begin{eqnarray}
\Delta V^2/C^2_s & = &  \frac{\mu}{\gamma A}+(Q \sin i \tan i \sin
2 \theta)^2
+\sin^2 i(M_{RR} \cos^2\theta-2 M_{R\theta} \cos \theta \sin \theta\\
\nonumber & + & M_{\theta \theta} \sin^2 \theta)+2 \sin i \cos i (M_{Z R} 
\cos \theta - M_{Z \theta} \sin \theta)+ \cos^2 i M_{ZZ}\\
\end{eqnarray}
where $i$ is the orbital inclination, $\mu$ is the mean molecular weight, 
$\gamma=5/3$, and $A$ is the atomic weight. The shear parameter $Q$ is 
given by
\begin{equation}
Q=\frac{3}{4}\frac{H}{R}\frac{V_{\rm Kep}}{C_s}\frac{\Delta Z}{2 H}
\end{equation}
and, for simplicity, is set to $3/4$, the maximum value for an optically 
thin accretion disk.

The Mach matrix is given by the correlations between components of 
turbulence in different cylindrical coordinates:
\begin{equation}
M_{ij}\equiv \frac{\langle \delta V_i \delta V_j \rangle}{C_s^2}
\end{equation}
where $C_s$ is the sound speed.

\clearpage

\begin{table}
\begin{tabular}{llll}
Symbol & Adopted value & Meaning & Reference\\
\hline
$q$ & 0.58 & Binary mass ratio & Howarth \&\ Wilson 1983a\\
$M_{\rm ns}$ & 1.4 & Mass of neutron star (M$_\odot$) & \\
$R_{\rm inner}$ & 10$^6$ & Inner radius of accretion disk (cm) & Cheng, 
Vrtilek, \&\ Raymond 1995\\
$R_{\rm outer}$ & 2$\times10^{11}$ & Outer radius of accretion disk (cm) 
& CVR\\
$a$    & 0.5 & Albedo & CVR\\
$D$ & 6600 & Distance to system (pc) & Reynolds et al. 1997\\
E$_{\rm B-V}$ & 0.018 & Reddening parameter & Boroson et al. 2001\\
T$_*$ & 8100 & Polar temperature of normal star (K) & CVR\\
$\Delta \Psi$ & $0.8105$ & Disk precession phase offset & H\&W \\
 $\theta_{\rm d}$ & 4.87 & Disk thickness (degrees) & H\&W \\
$\alpha_{\rm d}$ & 28.72 & Disk tilt from orbital plane (degrees) & H\&W \\
$a_*$ & 6.35$\times 10^{11}$ & Orbital separation, centers of mass (cm) & 
H\&W \\ $i$ & 81.25 & Orbital inclination (degrees) & H\&W \\
\end{tabular}
\caption{Parameters describing the X-ray binary Hercules X-1}
\end{table}

\begin{table}
\begin{tabular}{lllcrl}
Root name & MJD (start) & MJD (end) & Orbital Phase & Exptime (s) 
& Comments\\
\hline
B0080101001  &  52069.32464 &  52069.36421 & 0.182562 & 3418 & \\
B0080101002  & 52069.39311 &  52069.43359 & 0.223105 & 3498 & \\
B0080101003   & 52069.46245 &  52069.50297 & 0.263900 & 3501 & \\
B0080101004   & 52069.53186 &  52069.57233 & 0.304712 & 3497 & \\
B0080101005   & 52069.60135 &  52069.64170 & 0.345547 & 3455 & Dip in 
O\,{\sc vi} (double-peak gap?)\\
B0080101006   & 52069.67566 &  52069.71106 & 0.387799 & 2800 & Dip in 
O\,{\sc vi}\\ 
B0080101007   & 52069.74947 &  52069.78043 & 0.429904 & 2675  & \\
B0080101008   & 52069.82233 &  52069.84980 & 0.471734 & 2374 & \\
B0080101009   & 52069.89492 &   52069.91917 & 0.513482 & 2095 & \\
B0080101010  & 52069.96775 &   52069.98854 & 0.555305 & 1796 & \\
B0080101011  & 52070.04000 &   52070.05792 & 0.596954 & 1548 & \\
B0080101012  & 52070.11058 &   52070.12730 & 0.638113 & 1442 & \\
B0080101013  & 52070.15631 &   52070.16500 & 0.662651 & 644 & \\
B0080101014  & 52070.18235 &   52070.19666 & 0.679620 & 1231 & \\
B0080101015  & 52070.22613 &   52070.26604 & 0.712899 & 3447 & \\
B0080101016  & 52070.29521 &   52070.33540 & 0.753612 & 3472 & \\
B0080101017  & 52070.36448 &   52070.40475 & 0.794380 & 3479 & 
O\,{\sc vi}$\lambda1032$ double-peaked\\ 
B0080101018  & 52070.43388 &   52070.47413 & 0.835192 & 3478 & 
O\,{\sc vi}$\lambda1032$ double-peaked\\ 
B0080101019  & 52070.50334 &   52070.54349 & 0.876021 & 3469 & 
Lines anomalously bright\\
B0080101020  & 52070.57267 &   52070.61286 & 0.916809 & 3434 & \\
B0080101021  & 52070.64515 &   52070.68223 & 0.958526 & 3204 & \\
B0080101022  & 52070.71889 &   52070.75160 & 0.000611 & 2811 & Mid-Eclipse\\
B0080101023  & 52070.79200 &   52070.82097 & 0.042515 & 2503 & \\
B0080101024  &   52070.86463 &   52070.88778 & 0.083521 & 2000 & \\
\end{tabular}
\caption{A log of the {\it FUSE} Her X-1 observations.}
\end{table}

\begin{table}
\begin{tabular}{rrlllll}
Exposure & Orbital Phase & Duration (s) & $\dot{M}$ (Kurucz)  & 
$\chi^2$ (Kurucz) & $\dot{M}$ ({\it FUSE}) & $\chi_{\nu}^2$ ({\it FUSE})\\
         &           &      & ($-\log_{10}$ M$_\odot$) & & ($-\log_{10}$ 
M$_\odot$) \\ \hline
1  & 0.182 & 3418 & 8.04 & 2.7 & 8.06 & 2.7\\
2  & 0.223 & 3498 & 8.07 & 2.9 & 8.10 & 3.0\\
3   & 0.264 & 3501 & 8.09 & 3.1 & 8.19 & 3.2\\
4   & 0.305 & 3497 & 8.16 & 2.9 & 8.26 & 3.0\\
5   & 0.346 & 3486 & 8.21 & 2.7 & 8.32 & 2.9\\
6   & 0.388 & 3059 & 8.24 & 2.5 & 8.33 & 2.7\\
7   & 0.430 & 2675 & 8.29 & 2.4 & 8.38 & 2.8\\
8   & 0.472 & 2374 & 8.20 & 2.3 & 8.30 & 2.5\\
9   & 0.513 & 2095 & 8.07 & 2.6 & 8.18 & 2.7\\
10  & 0.555 & 1796 & 8.08 & 2.4 & 8.17 & 2.7\\
11  & 0.597 & 1548 & 8.13 & 2.5 & 8.21 & 2.6\\
12  & 0.638 & 1444 & 8.20 & 2.2 & 8.27 & 2.3\\
13  & 0.663 & 751 & 8.22 & 1.2 & 8.30 & 1.3\\
14  & 0.680 & 1236 & 8.19 & 1.6 & 8.27 & 1.6\\
15  & 0.713 & 3448 & 8.17 & 2.5 & 8.27 & 2.6\\
16  & 0.754 & 3473 & 8.20 & 2.5 & 8.22 & 2.6\\
17  & 0.794 & 3480 & 8.18 & 2.3 & 8.19 & 2.3\\
18  & 0.835 & 3478 & 8.22 & 2.2 & 8.22 & 2.2\\
19  & 0.876 & 3469 & 8.17 & 2.2 & 8.18 & 2.2\\
20  & 0.917 & 3472 & 8.29 & 1.7 & 8.28 & 1.7\\
21  & 0.958 & 3204 & NA &   NA  &  NA &  NA\\
22  & 0.001 & 2826 & NA &   NA  &  NA &  NA\\
23  & 0.042 & 2503 & NA &   NA  &  NA &  NA\\
24  & 0.084 & 2000 & 8.10 & 1.7 & 8.12 & 1.7\\
\end{tabular}
\caption{Fits to the Her~X-1 continuum as observed with {\it FUSE}.
$\dot{M}$ (Kurucz) and $\chi^2$ (Kurucz) give the mass accretion rate
and goodness of fit parameter for the model fit that uses Kurucz spectra 
while $\dot{M}$ (FUSE) and $\chi^2$ (FUSE) give those parameters for fits 
using a library of {\it FUSE} spectra.} \end{table}

\begin{table}
\begin{tabular}{llllll}
Ion & Wavelength (\AA) & $r_{\rm star}$ & $r_{\rm disk}$ & $r_{\rm 
continuum}$ & Comments\\
\hline
S\,{\sc vi} & 933.4 & 0.88 & 0.38 & 0.67 & \\
S\,{\sc vi} & 944.5 & 0.88 & 0.54 & 0.85 & \\
C\,{\sc iii} & 977 & 0.90 & 0.45 & 0.76 & Affected by strong ISM line\\
N\,{\sc iii} & 992 & 0.80 & 0.62 & 0.91 & Near strong airglow lines\\
O\,{\sc vi} & 1031.9 & 0.74 & 0.75 & 0.95 & \\
O\,{\sc vi} & 1037.6 & 0.82 & 0.71 & 0.98 & \\
S\,{\sc iv} & 1073 & 0.75 & 0.63 & 0.88 & \\
P\,{\sc v} & 1128 & 0.75 & 0.58 & 0.86 & \\
C\,{\sc iii} & 1175 & 0.82 & 0.65 & 0.95 \\
\end{tabular}
\caption{Correlation coefficients between the light curves of emission 
lines and continuum components as separated by our model 
(\S5).} 
\end{table} 

\begin{table}
\begin{tabular}{crrlrr}
Star & Spectral Type & T$_{\rm eff}$ (K) & FUSE Rootname & E(B-V) & V\\
\hline\\
HD~146813 & B\,1.5 V & 24000 & P1014901 & 0.02 & 9.06\\
HD~74662 & B\,3 V & 18000 &    A1290201 & 0.09 & 8.87 \\
(Interpolated) & B\,4 V & 16500 &  & \\
HD~92288 & B\,6 V & 14000 &    Z9012801 & 0.05 & 7.9\\
(HD~21672,HD~92536) & B\,8 V & 11500 &          (Z9011401,Z9012901) & 
(0.09,0.04) &
	(6.63,6.32)\\
HD~149630 & B\,9 V & 10800 &   B0910101    & 0.04 & 4.2   \\
(HD~109573,HD~181296) & A\,0 V & 10000 &   (B0910401,P2500101)    & 
(0.0,0.01) & 
		(5.78,5.03)  \\ 
HD~31647 & A\,1 V & 9300 &     C0380901    & 0.0 & 4.989 \\
HD~115892 & A\,2 V & 9050 &    A0410505    & 0.0 & 2.75 \\
HD~43940 & A\,3 V & 8850 &     C0380101    & 0.0 & 5.87 \\
HD~11636 & A\,5 V & 8500 &     A0410101    & 0.0 & 2.64 \\
(Interpolated) & A\,6 V & 8350 & \\
HD~187642 & A\,7 V & 8050 &    D0990101    & 0.0 & 0.77 \\
\end{tabular}
\caption{FUSE spectra used to form a library of stellar continua versus 
effective temperature in the far UV.}
\end{table}

\clearpage

\setcounter{figure}{0}


\begin{figure}
\includegraphics[height=0.4\textheight]{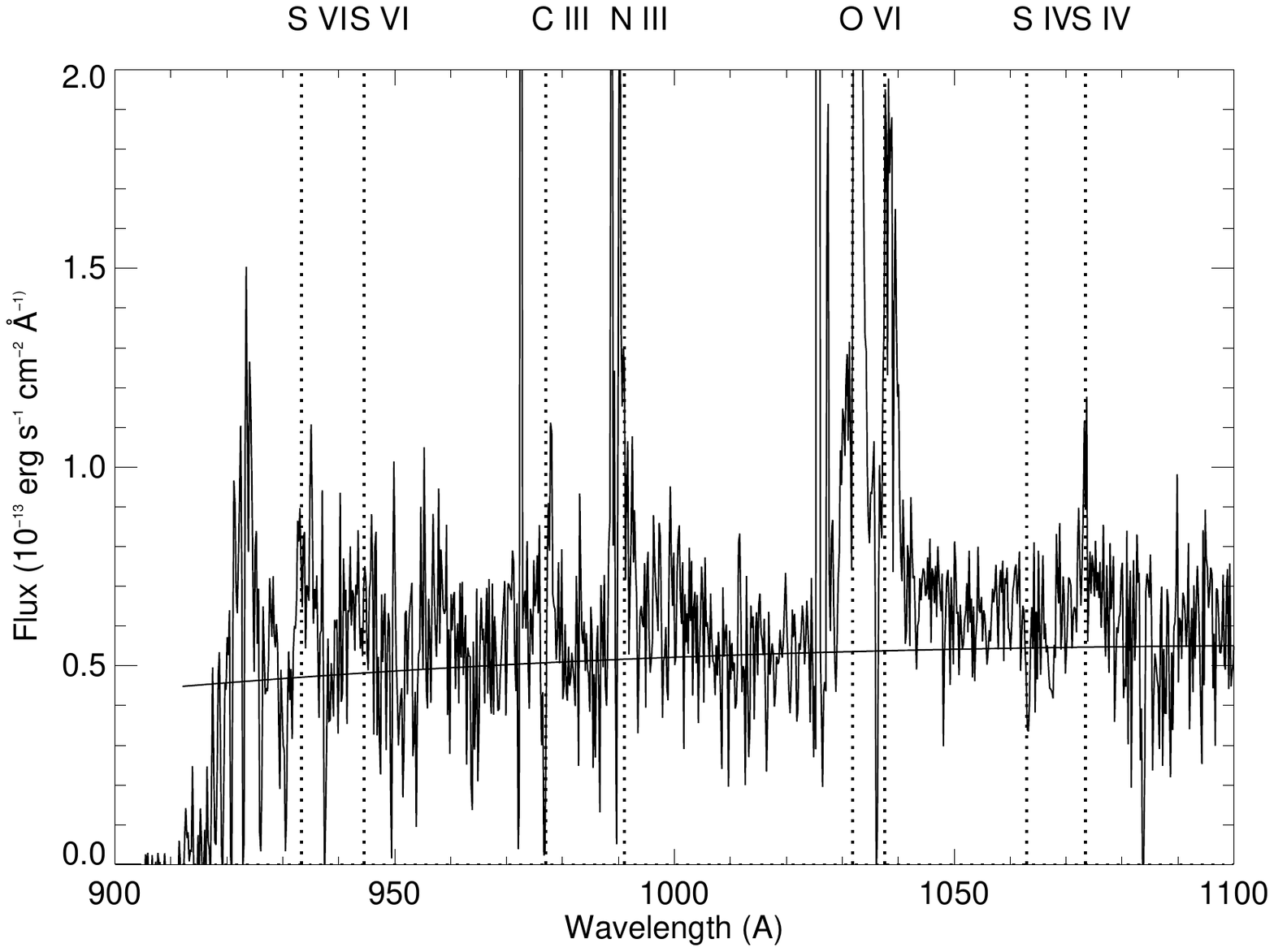}
\includegraphics[height=0.4\textheight]{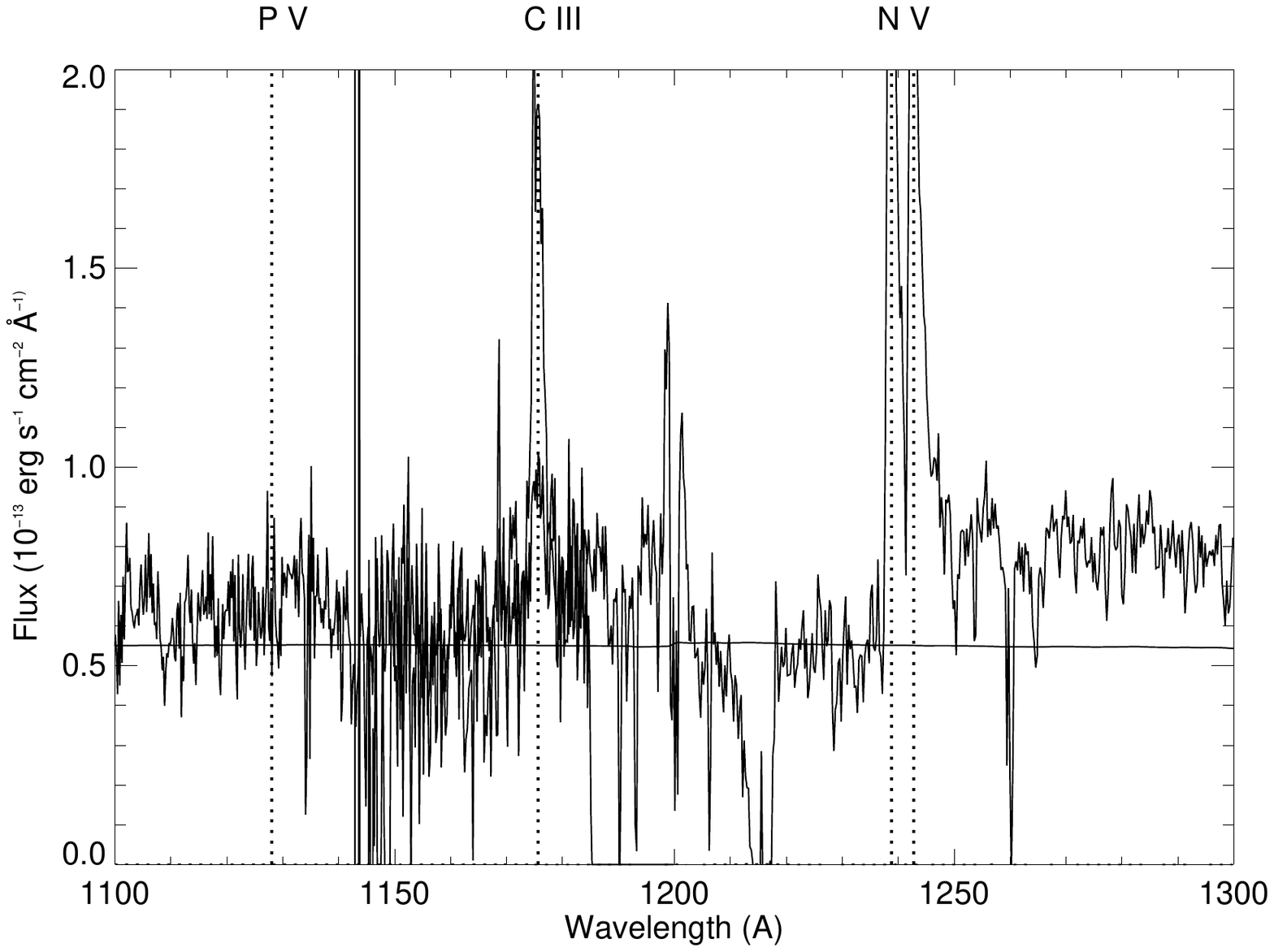}
\end{figure}

\newpage
\setcounter{figure}{0}
\begin{figure}
\includegraphics[height=0.4\textheight]{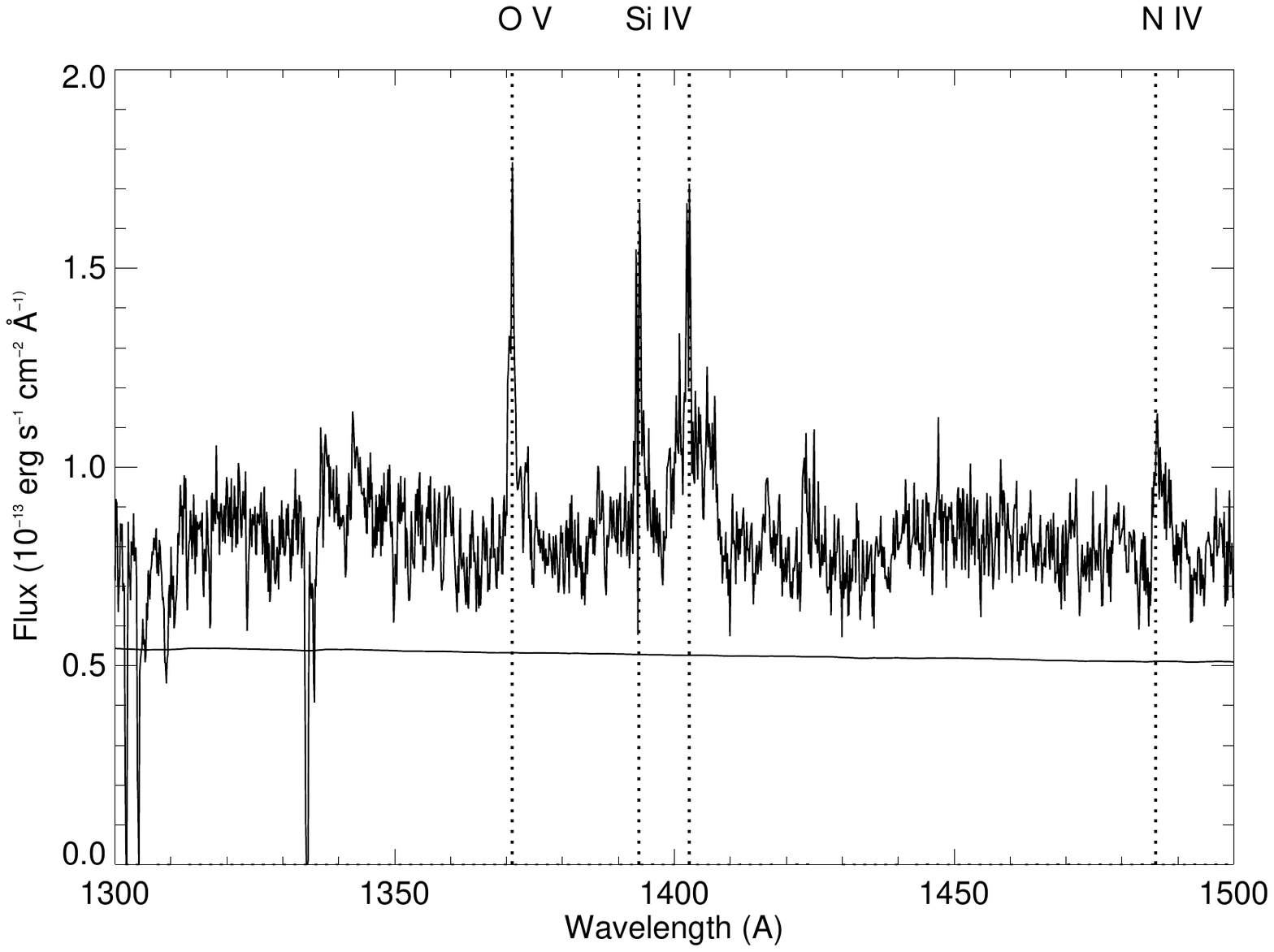}
\includegraphics[height=0.4\textheight]{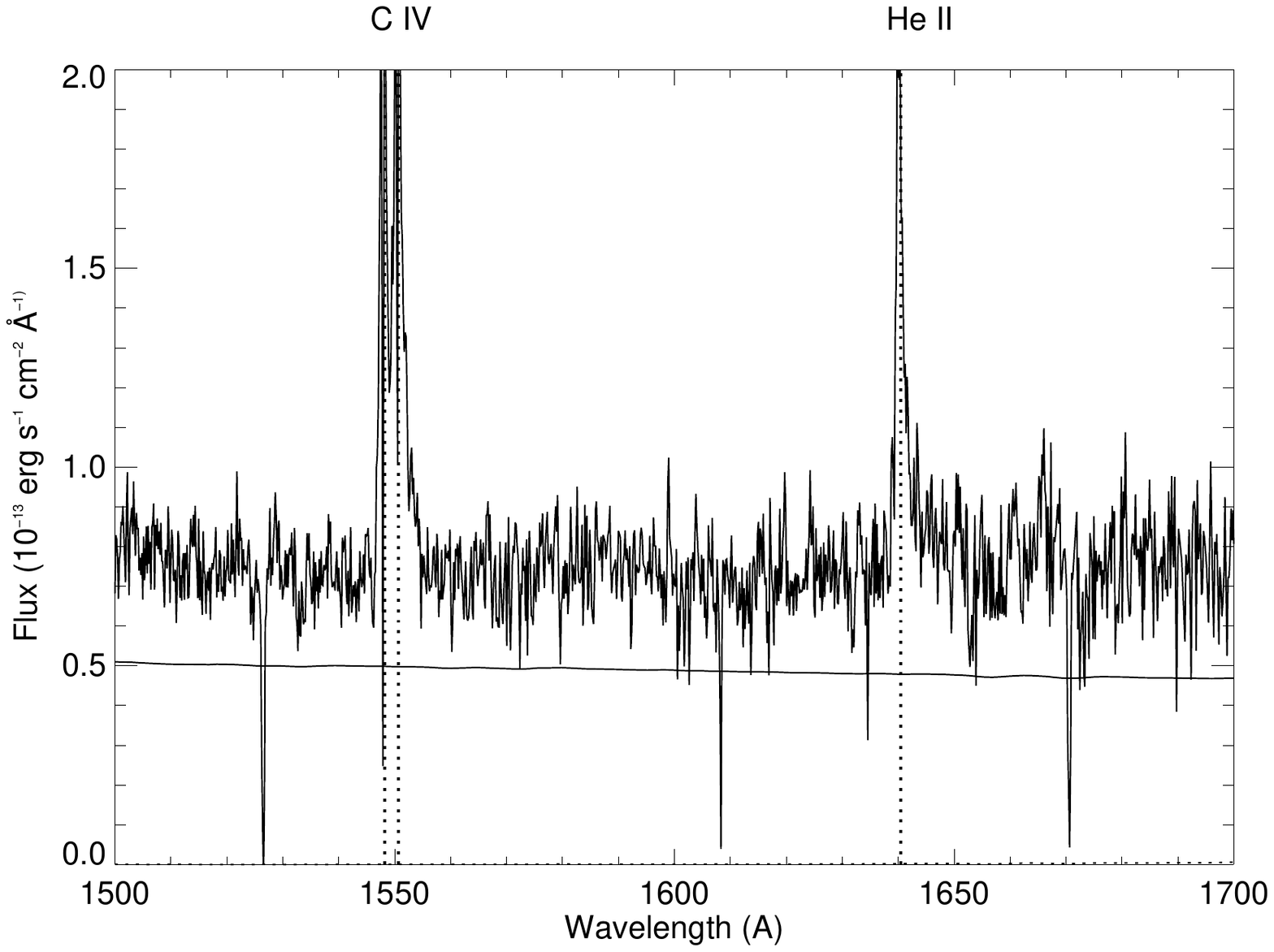}
\caption{The mean Her X-1 far UV spectrum observed with FUSE ($<1180$\AA) 
and the {\it HST} STIS ($>1150$\AA) near orbital phases when the emission is 
expected to be dominated by the accretion disk. Prominent 
emission lines are indicated by dotted vertical lines. 
We show he average of our continuum models using actual FUSE spectra of hot 
stars together with blackbodies. The component of the model resulting 
from the X-ray heated face of the normal star is indicated by a dotted 
curve.}
\end{figure}
\newpage


\newpage
\setcounter{figure}{1}
\begin{figure}
\includegraphics[height=0.4\textheight]{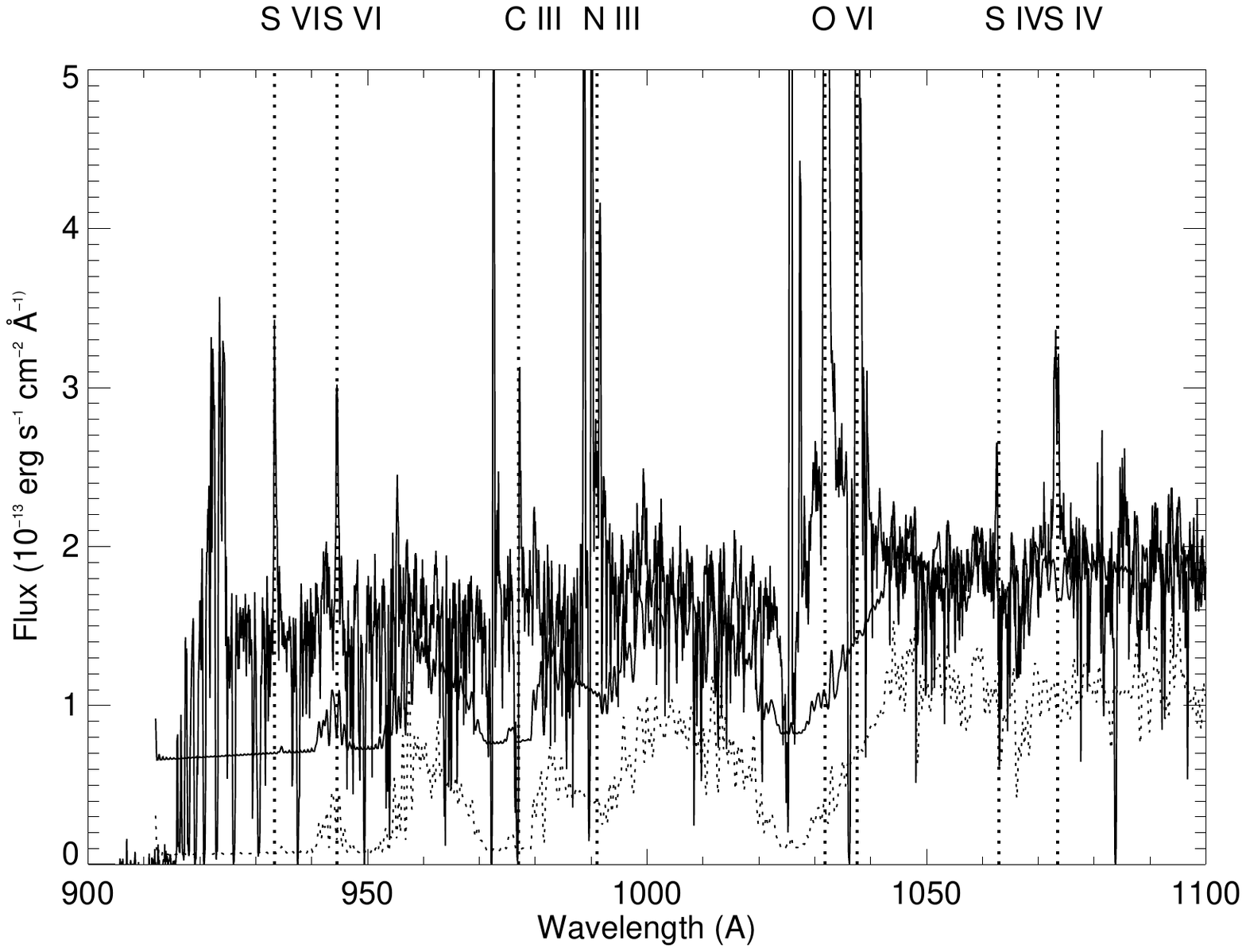}
\includegraphics[height=0.4\textheight]{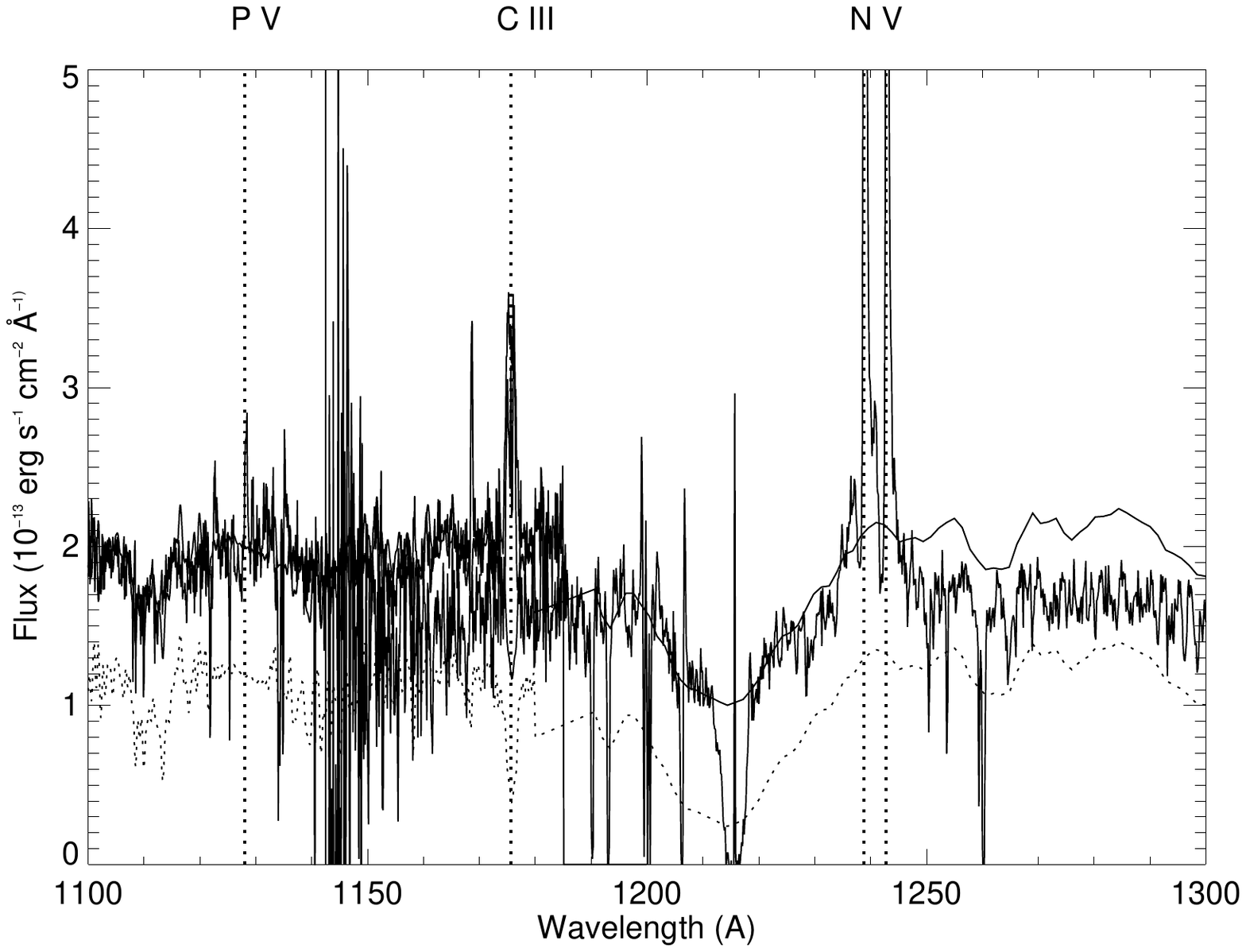}
\end{figure}
\newpage
\setcounter{figure}{1}
\begin{figure}
\includegraphics[height=0.4\textheight]{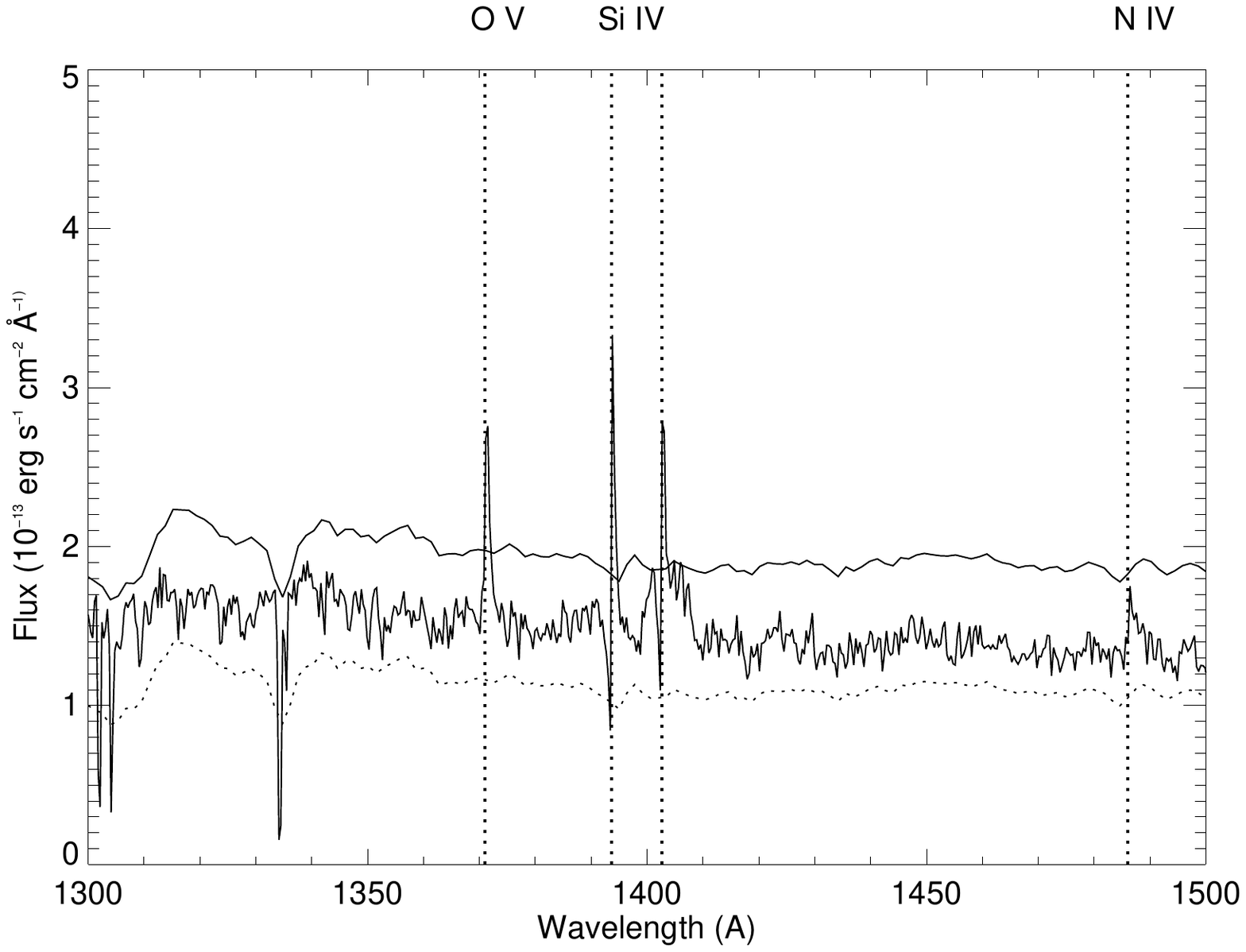}
\includegraphics[height=0.4\textheight]{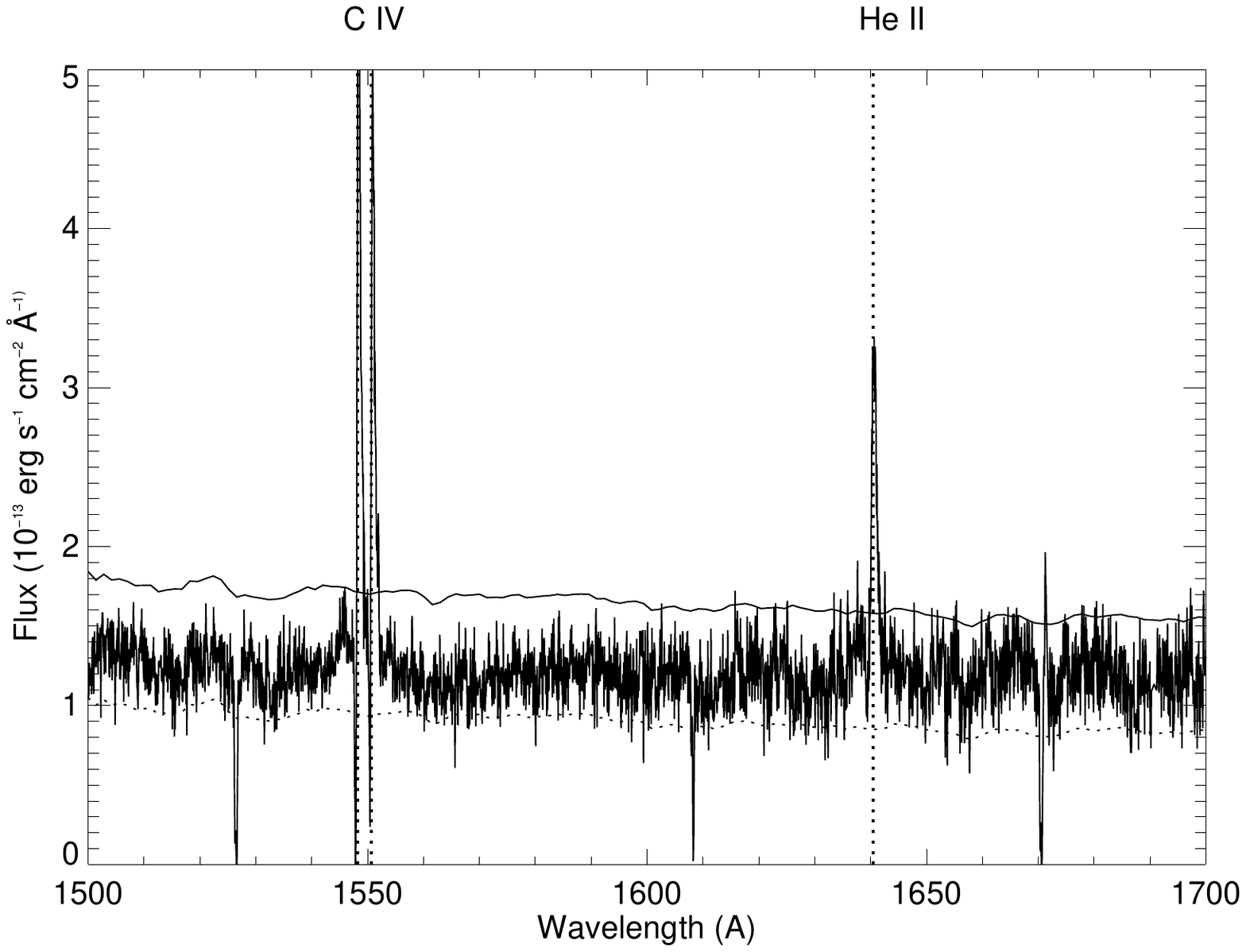}
\caption{The mean Her X-1 far UV spectrum observed with FUSE ($<1180$\AA) 
and the {\it HST} STIS ($>1150$\AA) near orbital phases when the emission is 
expected to be dominated by the X-ray heated face of the normal star.
The spectrum is plotted as in Figure~1. The lines and continuum are 
brighter at these phases, and the lines are narrower.}
\end{figure}
\newpage

\clearpage

\begin{figure}
\includegraphics[width=0.9\textwidth]{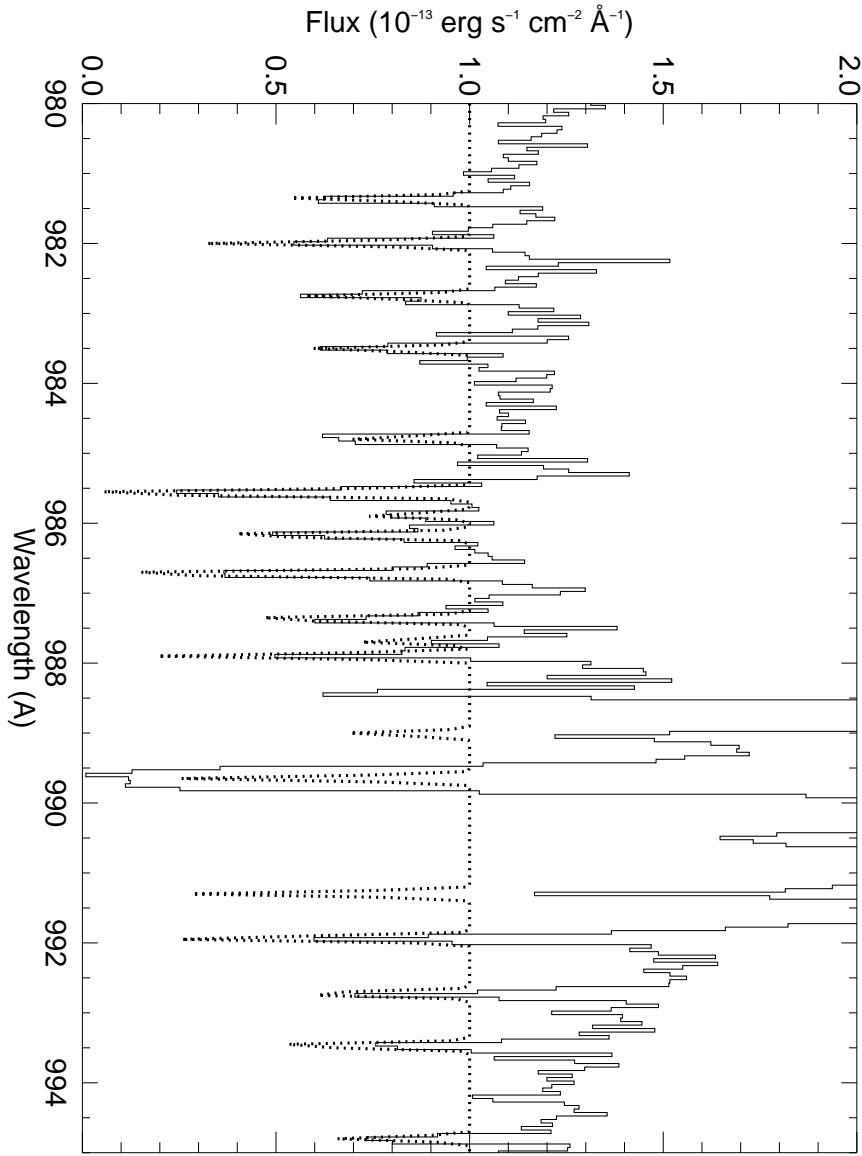}
\caption{A sample of the H$_2$ absorption lines in the spectrum of 
Her~X-1, and a model (dotted lines) with flat normalization, of absorption 
from rotational levels $J=0$ through $J=3$.}
\end{figure}

\setcounter{figure}{3}
\begin{figure}
\includegraphics[height=0.25\textheight]{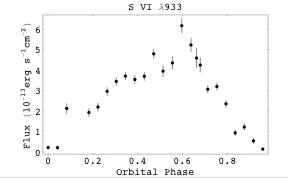}
\includegraphics[height=0.25\textheight]{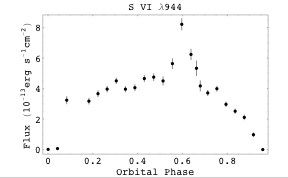}
\includegraphics[height=0.25\textheight]{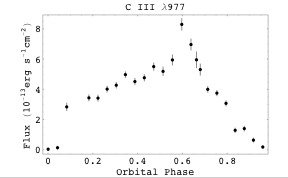}
\includegraphics[height=0.25\textheight]{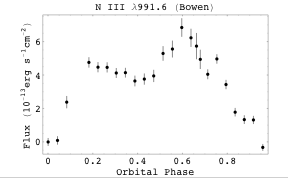}
\end{figure}
\setcounter{figure}{3}
\begin{figure}
\includegraphics[height=0.25\textheight]{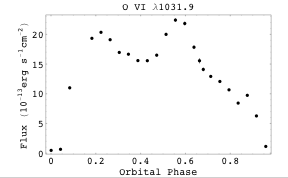}
\includegraphics[height=0.25\textheight]{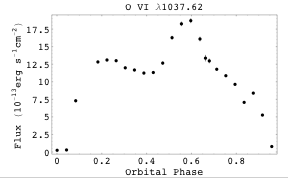}
\includegraphics[height=0.25\textheight]{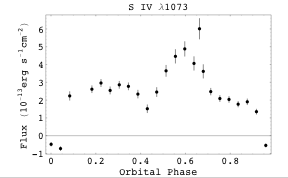}
\includegraphics[height=0.25\textheight]{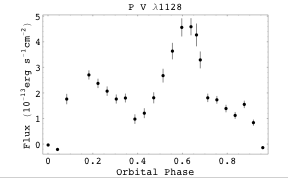}
\includegraphics[height=0.25\textheight]{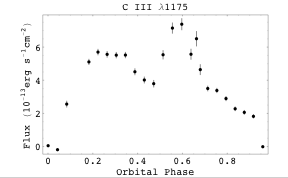}
\caption{Light curves of prominent lines in the FUSE spectrum of 
Hercules~X-1. A model of the continuum flux has been subtracted. Errors 
at the $1\sigma$ level are shown. Emission lines shown: (a) 
S\,{\sc vi}$\lambda933.4$, (b) S\,{\sc vi}$\lambda944.5$, 
(c) C\,{\sc iii}$\lambda 977$, (d) N\,{\sc iii}$\lambda991$, (e) O\,{\sc 
vi}$\lambda1031.9$, 
(f) O\,{\sc vi}$\lambda 1037.6$, (g) S\,{\sc iv}$\lambda1073$, (h) P\,{\sc 
v}$\lambda 1128$, (i) C\,{\sc iii}$\lambda 1176$}
\end{figure}

\newpage

\setcounter{figure}{4}
\begin{figure}
\epsscale{1.0}
\includegraphics[width=1.0\textwidth]{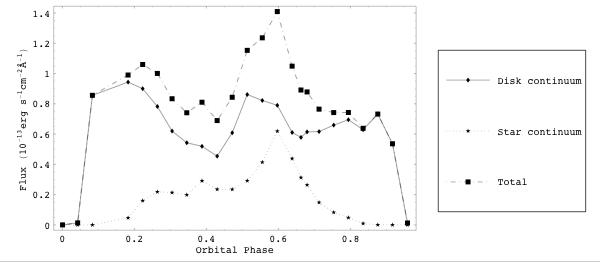}
\caption{Flux versus orbital phase $\phi$ in our model of the continuum 
flux in the 
wavelength range of the O\,{\sc vi} doublet. The continuum is separated 
into disk and stellar components.}
\end{figure}

\newpage

\setcounter{figure}{5}
\begin{figure}
\includegraphics[width=1.25\textwidth]{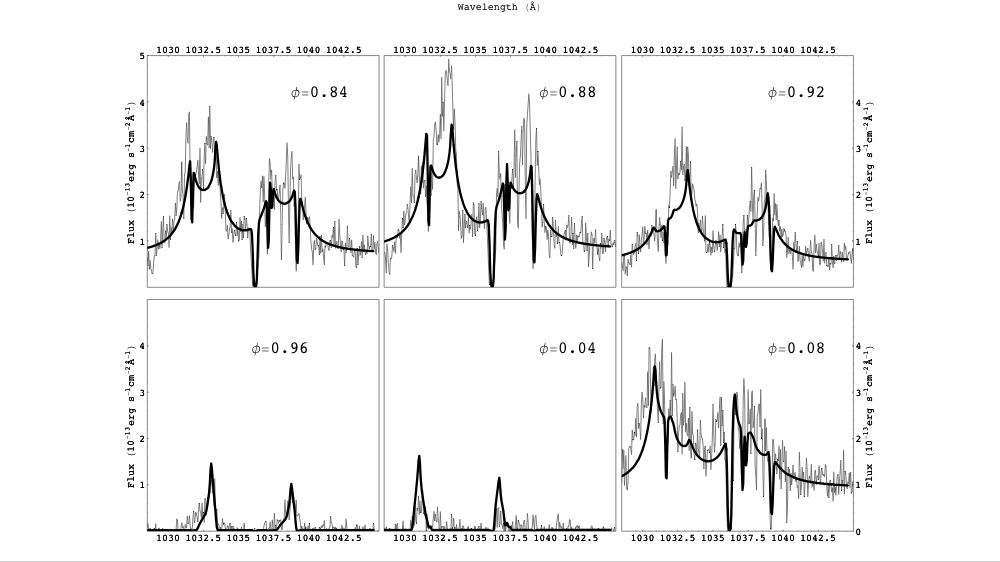}
\caption{Fits to the O\,{\sc vi} lines at phases when the accretion disk is 
partially eclipsed.}
\end{figure}

\newpage

\clearpage
\setcounter{figure}{6}
\begin{figure}
\includegraphics[width=0.8\textwidth]{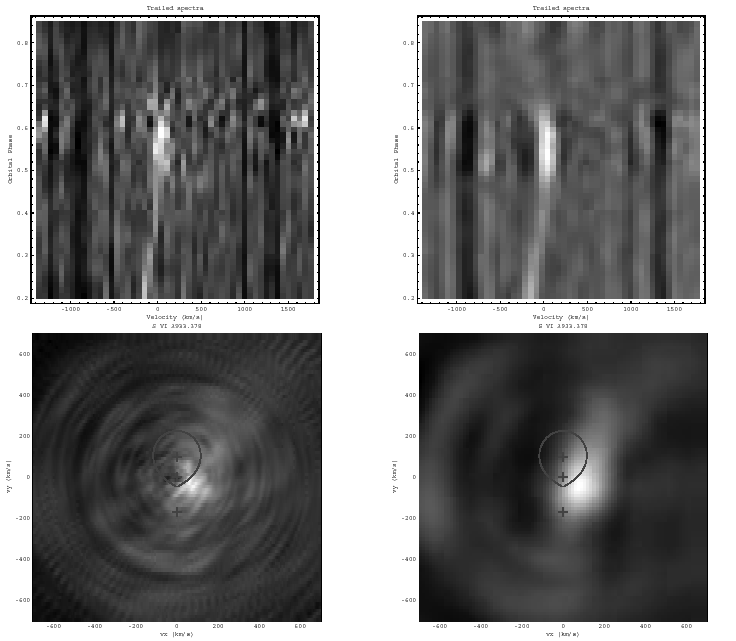}
\includegraphics[width=0.8\textwidth]{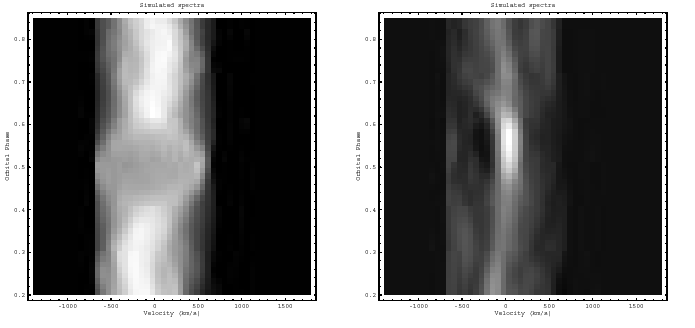}
\caption{A Fourier-filtered back-projected Doppler tomogram using the 
FUSE spectra of S\,{\sc vi}$\lambda 933$ from $\phi=0.2$ to $\phi=0.85$. 
The grayscale runs from the minimum to the maximum. 
Top: the trailed spectrogram (left) and filtered trailed spectrogram (right).
Middle: the back-projected tomogram (left) and filtered tomogram (right).
Bottom: the trailed spectrogram inverted from the back-projected 
tomogram (left) and from the filtered tomogram (right).}
\end{figure}

\clearpage

\begin{figure}
\includegraphics[width=0.95\textwidth]{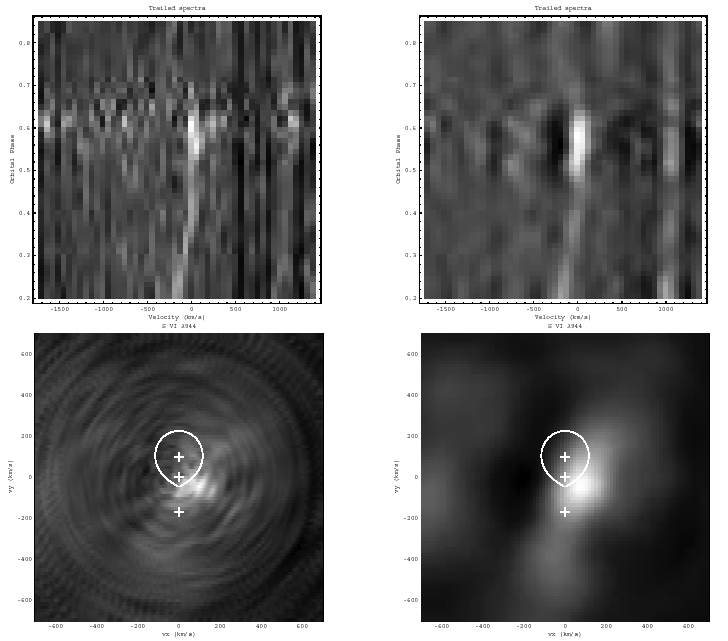}
\includegraphics[width=0.95\textwidth]{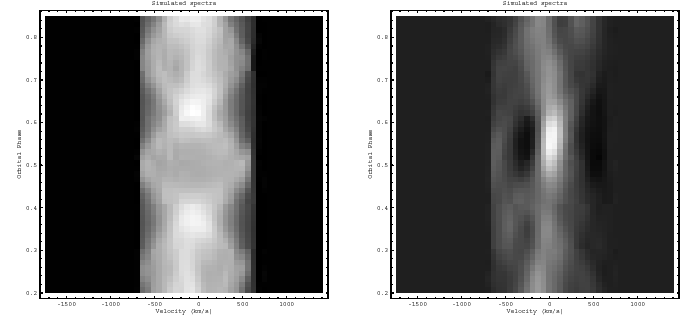}
\caption{A Fourier-filtered back-projected Doppler tomogram using the 
FUSE spectra of S\,{\sc vi}$\lambda 944$ presented as in Figure~7.}
\end{figure}

\begin{figure}
\includegraphics[width=0.95\textwidth]{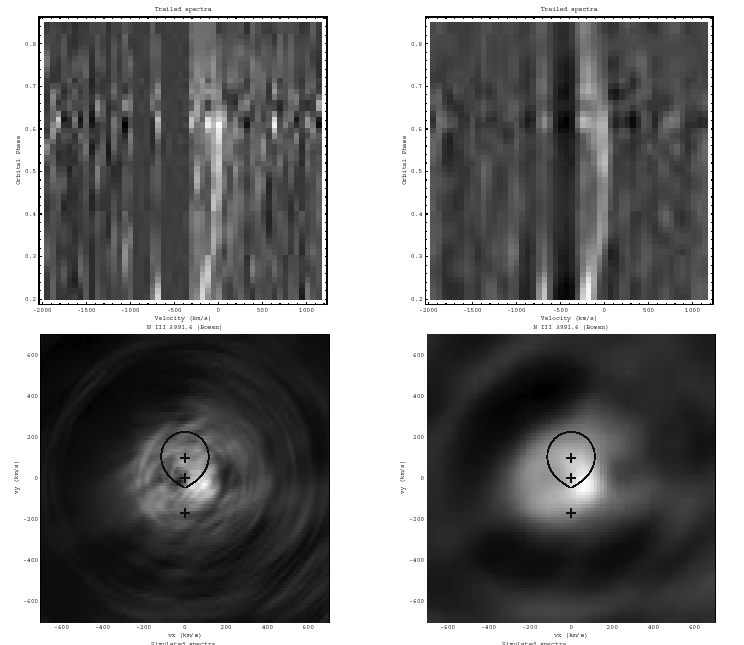}
\includegraphics[width=0.95\textwidth]{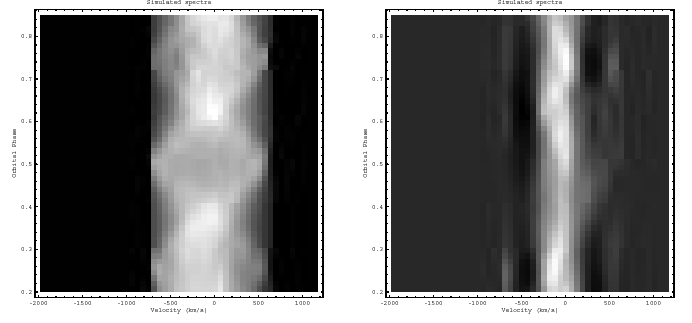}
\caption{A Fourier-filtered back-projected Doppler tomogram using the 
FUSE spectra of N\,{\sc iii}$\lambda 991.6$ (Bowen process) presented as in 
Figure~7.} \end{figure}

\clearpage

\begin{figure}
\includegraphics[width=0.95\textwidth]{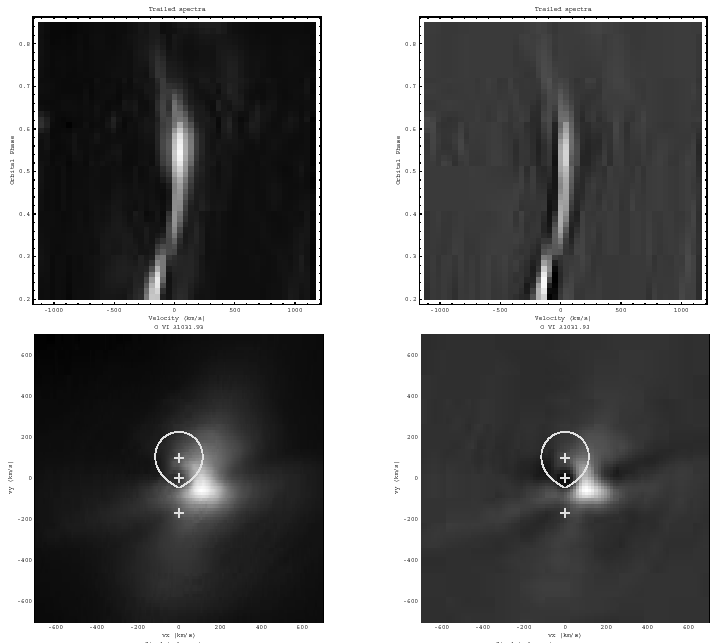}
\includegraphics[width=0.95\textwidth]{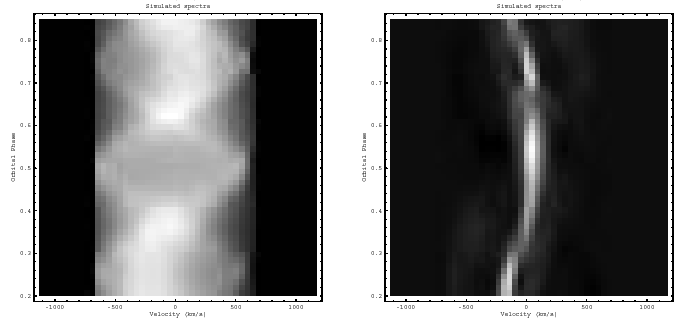}
\caption{A Fourier-filtered back-projected Doppler tomogram using the 
FUSE spectra of O\,{\sc vi}$\lambda 1031.9$ presented as in Figure~7.}
\end{figure}

\clearpage

\begin{figure}
\includegraphics[width=0.95\textwidth]{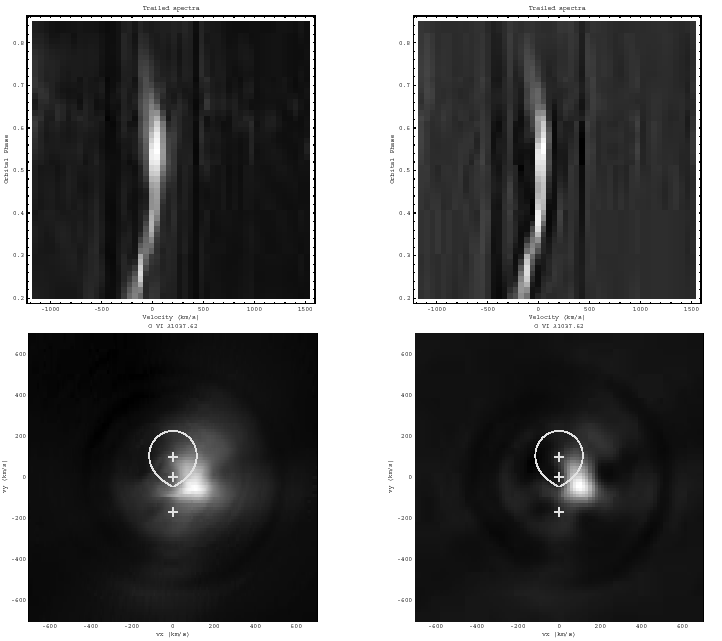}
\includegraphics[width=0.95\textwidth]{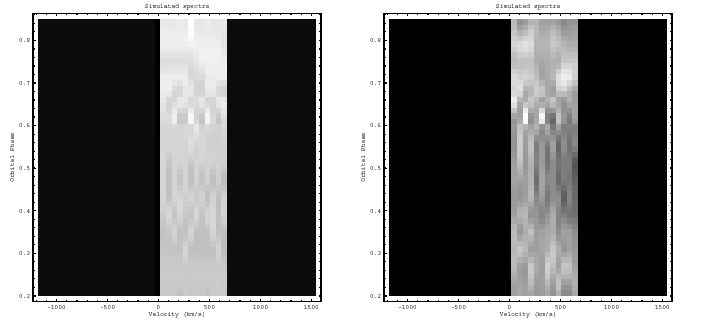}
\caption{A Fourier-filtered back-projected Doppler tomogram using the 
FUSE spectra of O\,{\sc vi}$\lambda 1037.6$ presented as in Figure~7.}
\end{figure}

\clearpage

\begin{figure}
\includegraphics[width=0.95\textwidth]{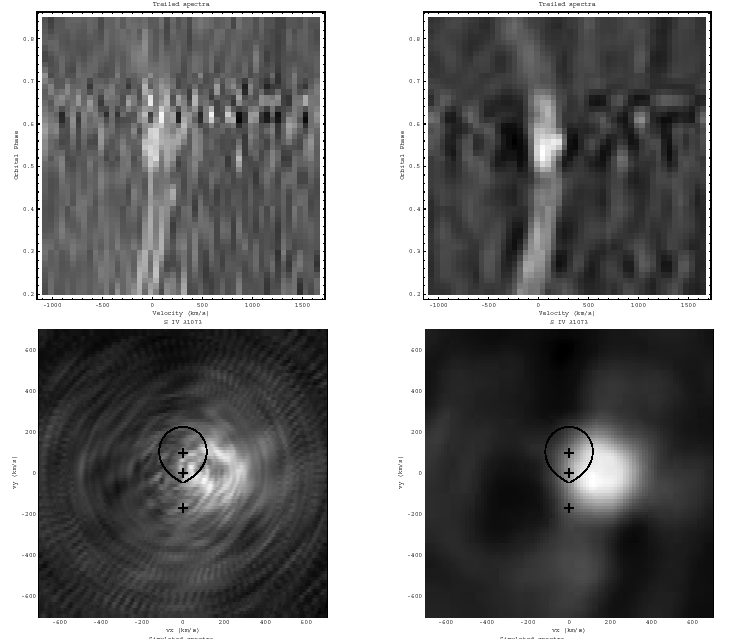}
\includegraphics[width=0.95\textwidth]{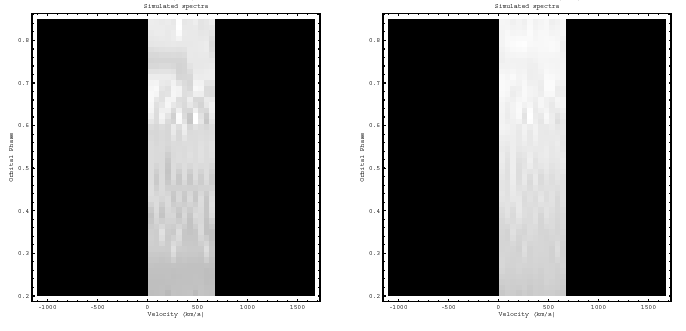}
\caption{A Fourier-filtered back-projected Doppler tomogram using the 
FUSE spectra of S\,{\sc iv}$\lambda 1073$ presented as in Figure~7.}
\end{figure}

\begin{figure}
\includegraphics[width=0.95\textwidth]{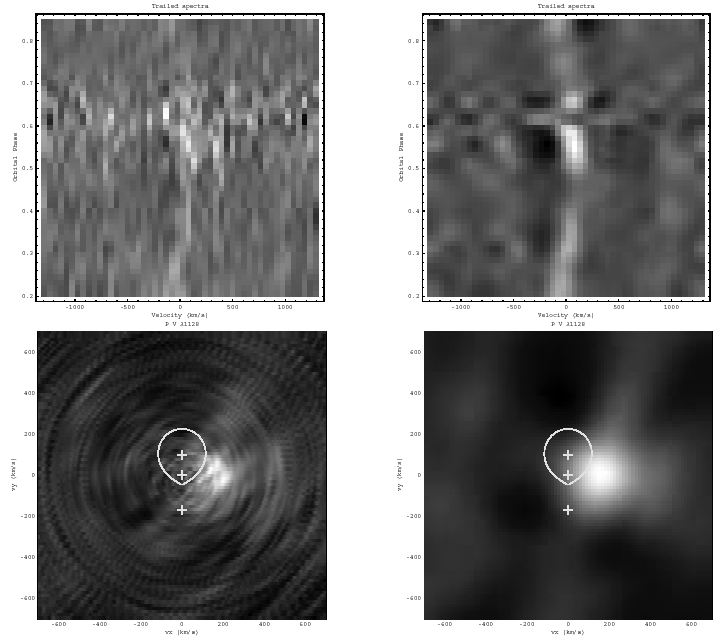}
\includegraphics[width=0.95\textwidth]{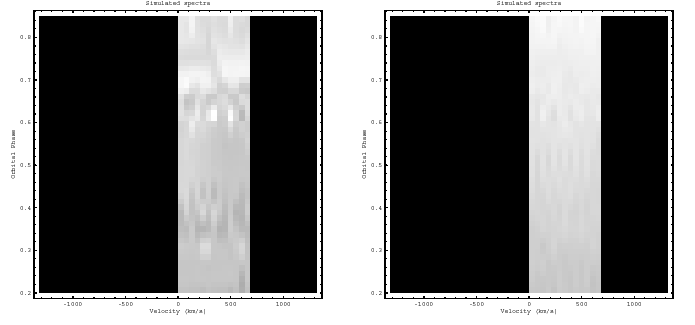}
\caption{A Fourier-filtered back-projected Doppler tomogram using the 
FUSE spectra of P\,{\sc v}$\lambda 1128$ presented as in Figure~7.}
\end{figure}
\clearpage

\begin{figure}
\includegraphics[width=0.95\textwidth]{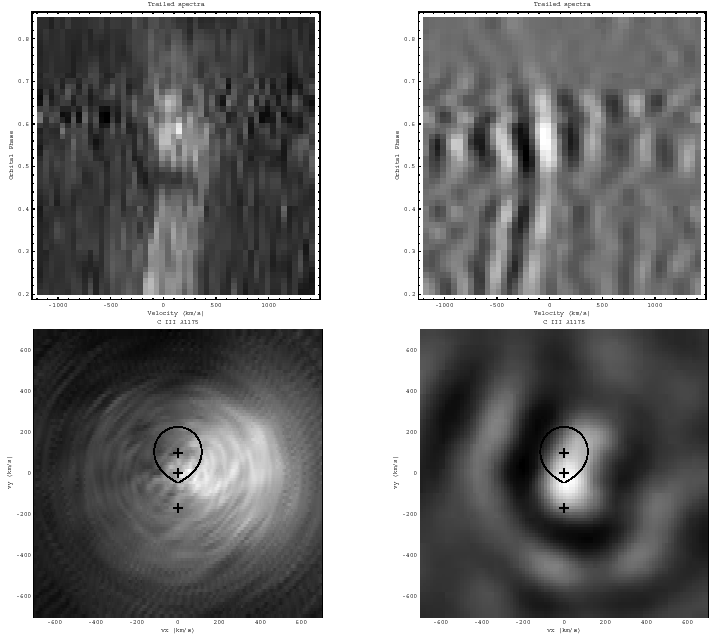}
\includegraphics[width=0.95\textwidth]{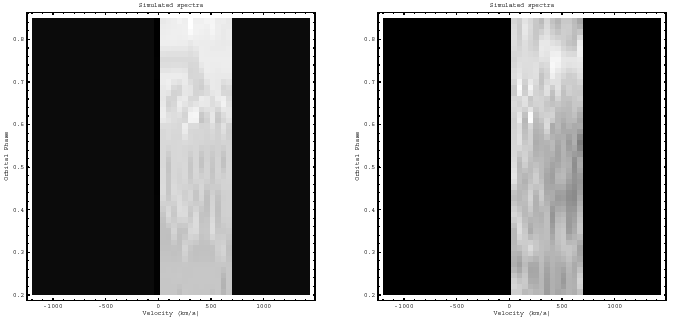}
\caption{A Fourier-filtered back-projected Doppler tomogram using the 
FUSE spectra of C\,{\sc iii}$\lambda 1176$ presented as in Figure~7.}
\end{figure}

\begin{figure}
\includegraphics[width=1.0\textwidth]{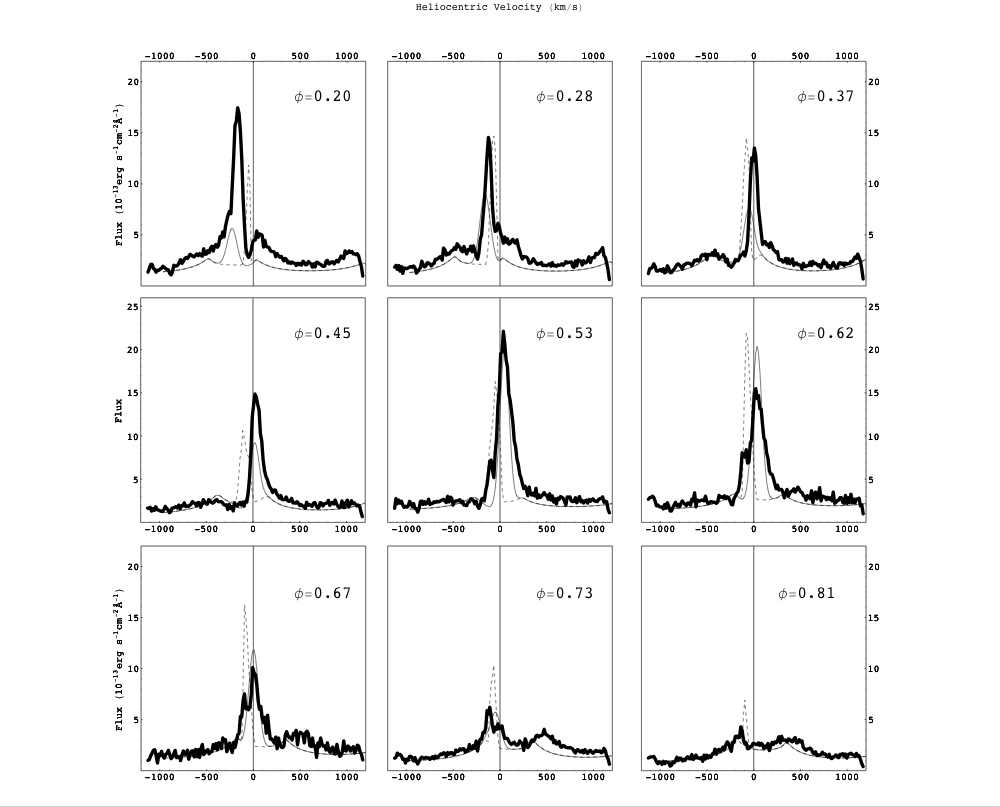}
\caption{The O\,{\sc vi}$\lambda1032$ emission line averaged over every 
two {\it FUSE} exposures (thick curve), a model of O\,{\sc vi} emission on 
the surface of the noncollapsed star (dashed), and an empirical model of 
Gaussian emission and absorption models. Both models include flux from a 
model of the accretion disk.}
\end{figure}

\begin{figure}
\includegraphics[width=0.85\textwidth]{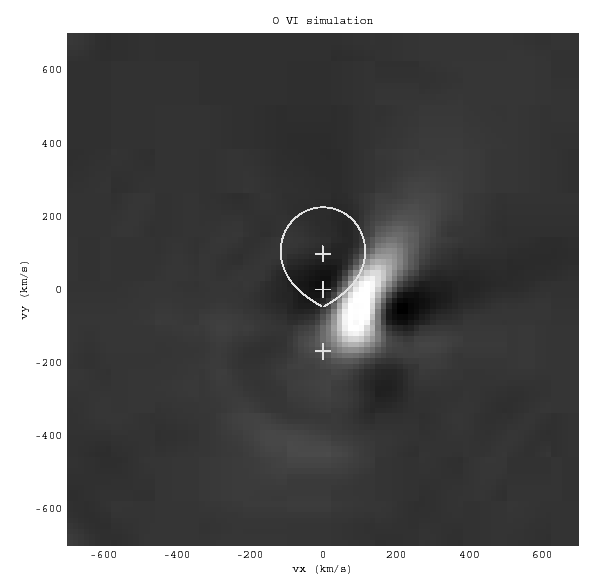}
\includegraphics[width=0.85\textwidth]{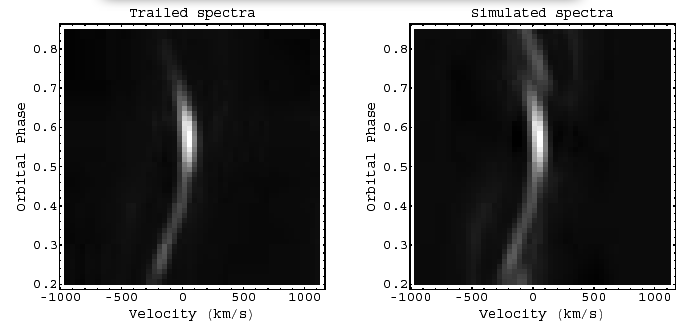}
\caption{A Doppler tomogram of a simple simulation of the 
O\,{\sc vi}$\lambda1032$ line as a sum of moving Gaussians and an accretion 
disk model, with counting statistics included.} 
\end{figure}

\begin{figure}
\includegraphics[width=1.0\textwidth]{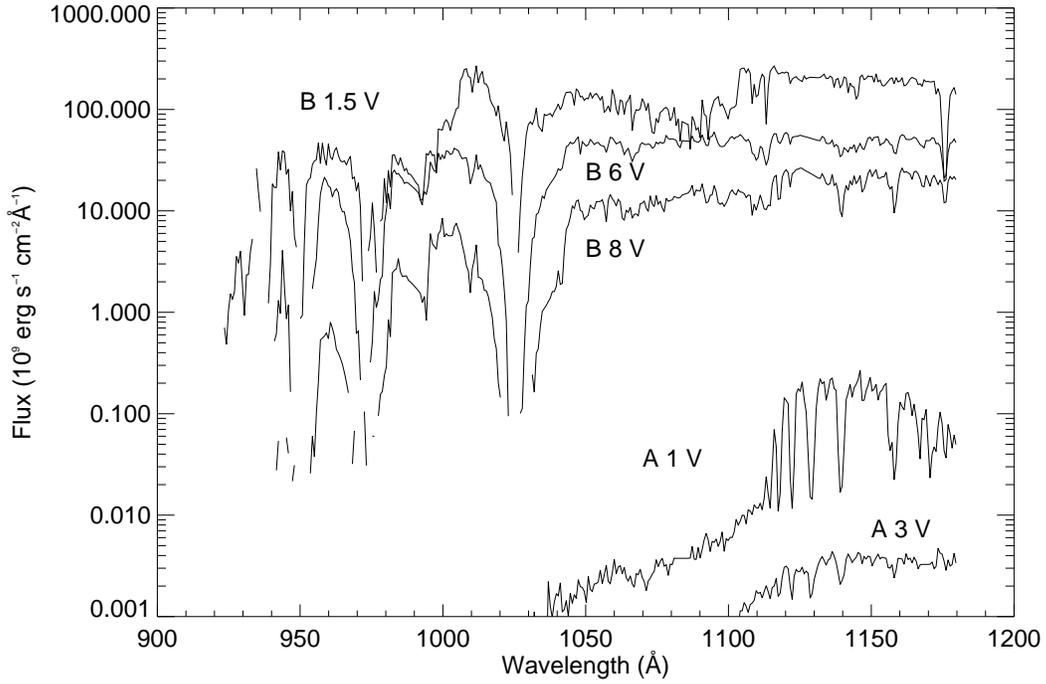}
\caption{Sample spectra, with spectral type labelled, in the library of 
{\it FUSE} spectra we use to complement our library of {\it IUE} and 
optical spectra. We use this library in the fitting of the Her~X-1 
continuum. All stars have their V band flux normalized to the 
same value.}
\end{figure}

\end{document}